\documentclass[preprint]{aastex63}

\newcommand{\eelg}{S$_1$ELG}
\newcommand{\selg}{S$_2$ELG}

\newcommand{\nzspec}{49} 
\newcommand{\nzspecOIII}{31} 
\newcommand{\nzEELG}{9} 
\newcommand{\nzSELG}{13} 
\newcommand{\nzSFG}{28}  

\newcommand{\zspecrange}{$2.091<z_{\rm spec}<4.751$}
\newcommand{\hizspecrange}{$3<z_{\rm spec}<3.8$}
\newcommand{\zphotrange}{$2.5<$\zphot$<4.0$}
\newcommand{\Qz}{$Q_{\rm z}$}

\newcommand{\spitzer}{{\it Spitzer}}

\newcommand{\fsjhk}{$J_1 J_2 J_3 H_s H_l K_s$}

\newcommand{\Ks}{$K_s$}
\newcommand{\uvj}{$UVJ$}

\newcommand{\umv}{$(U-V)$}
\newcommand{\vmj}{$(V-J)$}

\newcommand{\Luvir}{L$_{\rm UV+IR}$}
\newcommand{\mosel}{{\tt MOSEL}}
\newcommand{\prospector}{{\tt Prospector}}

\newcommand{\zfire}{{\tt ZFIRE}}
\newcommand{\zfourge}{{\tt ZFOURGE}}

\newcommand{\zphot}{$z_{\rm phot}$}
\newcommand{\zspec}{$z_{\rm spec}$}
\newcommand{\fesc}{$f_{\rm esc}$}
\newcommand{\ewobs}{EW$_{\rm obs}$}
\newcommand{\ewrest}{EW$_{\rm rest}$}

\newcommand{\sigmaz}{$\sigma_{z}$}
\newcommand{\intsigma}{$\sigma_{\rm int}$}

\newcommand{\kms}{km~s$^{-1}$}
\newcommand{\Halpha}{H$\alpha$}
\newcommand{\Hbeta}{H$\beta$}
\newcommand{\Lya}{Ly$\alpha$}
\newcommand{\HbOIII}{\Hbeta+\OIII}
\newcommand{\SII}{[\hbox{{\rm S}\kern 0.1em{\sc ii}}]}
\newcommand{\AlIII}{\hbox{{\rm Al}\kern 0.1em{\sc iii}}}
\newcommand{\NII}{[\hbox{{\rm N}\kern 0.1em{\sc ii}}]}
\newcommand{\OII}{[\hbox{{\rm O}\kern 0.1em{\sc ii}}]}
\newcommand{\OIII}{[\hbox{{\rm O}\kern 0.1em{\sc iii}}]}
\newcommand{\MgII}{\hbox{{\rm Mg}\kern 0.1em{\sc ii}}}
\newcommand{\MgI}{\hbox{{\rm Mg}\kern 0.1em{\sc i}}}
\newcommand{\FeII}{\hbox{{\rm Fe}\kern 0.1em{\sc ii}}}
\newcommand{\CIII}{\hbox{{\rm C}\kern 0.1em{\sc iii}}]}
\newcommand{\CIV}{\hbox{{\rm C}\kern 0.1em{\sc iv}}}

\newcommand{\Lsun}{${\rm L}_{\odot}$}
\newcommand{\Msun}{${\rm M}_{\odot}$}
\newcommand{\Msunyr}{${\rm M}_{\odot}$~yr$^{-1}$}
\newcommand{\Mstar}{${\rm M}_{\star}$}

\newcommand{\Mvir}{${\rm M}_{\rm dyn}$}
\newcommand{\logMstarMsun}{$\log({\rm M}_{\star}/{\rm M}_{\odot})$}
\newcommand{\ergu}{erg~s$^{-1}$~cm$^{-2}$~\AA$^{-1}$}
\newcommand{\mipsmu}{$24\mu$m}
\newcommand{\reff}{r$_{\rm eff}$}
\newcommand{\Avsed}{A$_{\rm V}$}

\newcommand{\Mgas}{${\rm M}_{\rm gas}$}
\newcommand{\fgas}{${\rm f}_{\rm gas}$}

\begin{document}

\title{\mosel: Strong \OIII5007\AA\ Emitting Galaxies at $(3<z<4)$ from the \zfourge\ Survey} 

\correspondingauthor{Kim-Vy H. Tran}
\email{kimvy.tran@gmail.com}

\author[0000-0001-9208-2143]{Kim-Vy H. Tran}
\affiliation{School of Physics, University of New South Wales, Kensington, Australia}
\affiliation{George P. and Cynthia W. Mitchell Institute for Fundamental Physics and Astronomy, Department of Physics \& Astronomy, Texas A\&M University, College Station, TX 77843, USA}
\affiliation{ARC Centre for Excellence in All-Sky Astrophysics in 3D (ASTRO 3D)}

\author[0000-0001-6003-0541]{Ben Forrest}
\affiliation{Department of Physics \& Astronomy, University of California, Riverside, CA 92521, USA}

\author[0000-0002-2250-8687]{Leo Y. Alcorn}
\affiliation{George P. and Cynthia W. Mitchell Institute for Fundamental Physics and Astronomy, Department of Physics \& Astronomy, Texas A\&M University, College Station, TX 77843, USA}
\affiliation{LSSTC Data Science Fellow}

\author[0000-0002-9211-3277]{Tiantian Yuan}
\affiliation{Swinburne University of Technology, Hawthorn, VIC 3122, Australia}
\affiliation{ARC Centre for Excellence in All-Sky Astrophysics in 3D (ASTRO 3D)}

\author[0000-0003-2804-0648]{Themiya Nanayakkara}
\affiliation{Leiden Observatory, Leiden University, P.O. Box 9513, NL 2300 RA Leiden, The Netherlands}

\author[0000-0003-1420-6037]{Jonathan Cohn}
\affiliation{George P. and Cynthia W. Mitchell Institute for Fundamental Physics and Astronomy, Department of Physics \& Astronomy, Texas A\&M University, College Station, TX 77843, USA}

\author[0000-0002-4653-8637]{Michael Cowley}
\affiliation{Centre for Astrophysics, University of Southern Queensland, West Street, Toowoomba, QLD 4350, Australia }
\affiliation{School of Chemistry, Physics and Mechanical Engineering, Queensland University of Technology, Brisbane, QLD 4001, Australia}
 
\author[0000-0002-3254-9044]{Karl Glazebrook}
\affiliation{Swinburne University of Technology, Hawthorn, VIC 3122, Australia}
\affiliation{ARC Centre for Excellence in All-Sky Astrophysics in 3D (ASTRO 3D)}

\author[0000-0002-8984-3666]{Anshu Gupta}
\affiliation{School of Physics, University of New South Wales, Kensington, Australia}
\affiliation{ARC Centre for Excellence in All-Sky Astrophysics in 3D (ASTRO 3D)}

\author[0000-0003-1362-9302]{Glenn G. Kacprzak}
\affiliation{Swinburne University of Technology, Hawthorn, VIC 3122, Australia}
\affiliation{ARC Centre for Excellence in All-Sky Astrophysics in 3D (ASTRO 3D)}

\author{Lisa Kewley}
\affiliation{Research School of Astronomy and Astrophysics, The Australian National University, Cotter Road, Weston Creek, ACT 2611, Australia}
\affiliation{ARC Centre for Excellence in All-Sky Astrophysics in 3D (ASTRO 3D)}

\author[0000-0002-2057-5376]{Ivo Labb\'e}
\affiliation{Swinburne University of Technology, Hawthorn, VIC 3122, Australia}

\author[0000-0001-7503-8482]{Casey Papovich}
\affiliation{George P. and Cynthia W. Mitchell Institute for Fundamental Physics and Astronomy, Department of Physics \& Astronomy, Texas A\&M University, College Station, TX 77843, USA}

\author[0000-0001-5185-9876]{Lee Spitler}
\affiliation{Department of Physics and Astronomy, Faculty of Science and Engineering, Macquarie University, Sydney, NSW 2109, Australia}
\affiliation{ARC Centre for Excellence in All-Sky Astrophysics in 3D (ASTRO 3D)}

\author[0000-0001-5937-4590]{Caroline M. S. Straatman}
\affiliation{Sterrenkundig Observatorium, Universiteit Gent, Krijgslaan 281 S9, 9000 Gent, Belgium}

\author[0000-0003-2008-1752]{Adam Tomczak}
\affiliation{Department of Physics, University of California, Davis, One Shields Ave., Davis, CA 95616}

\begin{abstract}

To understand how strong emission line galaxies (ELGs) contribute to the overall growth of galaxies and star formation history of the universe, we target Strong ELGs (SELGs) from the \zfourge\ imaging survey that have blended \HbOIII\ rest-frame equivalent widths of $>230$\AA\ and \zphotrange.  Using Keck/MOSFIRE, we measure \nzspec\ redshifts for galaxies brighter than \Ks$=25$ mag as part of our Multi-Object Spectroscopic Emission Line (\mosel) survey.  Our spectroscopic success rate is $\sim53$\% and \zphot\ uncertainty is \sigmaz=$[\Delta z/(1+z)]=0.0135$.  We  confirm 31 ELGs at \hizspecrange\ and show that Strong ELGs have spectroscopic rest-frame \OIII5007\AA\ equivalent widths of $100-500$\AA\ and tend to be lower mass systems [\logMstarMsun$\sim8.2-9.6$] compared to more typical star-forming galaxies.  The Strong ELGs lie $\sim0.9$~dex above the star-forming main-sequence at $z\sim3.5$ and have high inferred gas fractions of \fgas$\gtrsim60$\%, i.e. the inferred gas masses can easily fuel a starburst to double stellar masses within $\sim10-100$~Myr.  Combined with recent results using \zfourge, our analysis indicates that 1) strong \OIII5007\AA\ emission signals an early episode of intense stellar growth in low mass [M$_{\star}<0.1$M$^{\star}$] galaxies and 2) many, if not most, galaxies at $z>3$ go through this starburst phase.  If true, low-mass galaxies with strong \OIII5007\AA\ emission (\ewrest$>200$\AA) may be an increasingly important source of ionizing UV radiation at $z>3$.

\end{abstract}


\section{Introduction} 

Hierarchical formation predicts that massive galaxies like our own Milky Way grow through the merger and accretion of smaller systems \citep{peebles:70}, thus low-mass galaxies that are chemically pristine can provide insight into the early stages of galaxy formation.  Although low-mass galaxies are abundant, identifying the ones that are the least chemically evolved via emission lines is difficult due to their rare nature in the local universe.  In the past decade, dwarf galaxies with strong \OIII5007\AA\ emission at $z\lesssim0.3$ \citep{cardamone:09,izotov:11} have been identified using optical imaging where the large equivalent width of the emission line increases the broadband flux.  Valuable insight is gained by measuring, e.g. star formation conditions and ionizing escape fractions \citep{amorin:12,jaskot:13,izotov:16,bian:16,lofthouse:17,izotov:18}, in these local ``green pea'' galaxies.  For example, studies find that strong \Hbeta+\OIII5007\AA\ emission are ubiquitous in Lyman-break galaxies at $z\sim7$ \citep{smit:14}.

A number of studies have now identifed dwarf galaxies (\logMstarMsun$\lesssim9$) at $0<z<2$ with strong \OIII5007\AA\ emission that may bridge local ``green peas'' to primeval galaxies at $z>6$.  Slit-less near-infrared spectroscopy with the {\it Hubble Space Telescope} has revealed a population of dwarf galaxies up to $z\sim2$ with rest-frame \OIII5007\AA\ equivalent widths of \ewrest$>200$\AA\ \citep[e.g.][]{straughn:09,vanderwel:11,atek:11}.  Dedicated ground-based spectroscopic surveys also have identifed Strong Emission Line Galaxies (SELGs) up to $z\sim1$ \citep{amorin:15}.  These studies suggest that the number density of SELGs increases with redshift \citep[e.g.][]{maseda:18}.  However, quantifying the evolving number density of strong \OIII5007\AA\ emitting galaxies at $z>1$ requires near-infrared spectroscopy ($\lambda>1\mu$m), thus only a handful of systems have been confirmed at $z\gtrsim3$ \citep{debarros:16,nakajima:16,amorin:17}.

Once identified, the natural question then is how these strong \OIII5007\AA\ emitting galaxies fit into our existing picture of galaxy formation.  The increasing number of Strong ELGs combined with the brief duration of this intense starburst phase \citep[$\lesssim100$~Myr;][]{guo:16,ceverino:18} supports a model where galaxies grow through multiple intense starbursts.  For starburst systems at $z\gtrsim3$ with low metallicities, such an episode can signal the initial major growth spurt, i.e. the ELGs with the highest equivalent widths (\ewrest$\gtrsim1000$\AA) are ``first burst'' systems \citep{cohn:18}.  In combination with studies of, $e.g.$ Lyman-break galaxies at $z\sim7$ with strong \Hbeta+\OIII5007\AA\ emission  \citep{roberts-borsani:16}, we can use SELGs to test current galaxy formation models that capture the intricate interplay of physics on the sub-kpc scale with the integrated galaxy properties that can be measured at $z>2$ \citep{krumholz:17}.  

An increasing population of Strong ELGs with redshift also has important implications for cosmic reionization.  These vigorously star-forming galaxies have steep UV slopes ($\beta\lesssim-2$) and low metallicities \citep[$Z/Z_{\odot}\lesssim0.2$;][]{forrest:18,cohn:18}, i.e. the SELGs may be a tremendous source of UV photons.  By identifying the strong \OIII5007\AA\ emitting galaxies, we can then measure their Lyman-Continuum emission and escape fractions to infer if SELGs at $z>8$ can generate the UV photons needed to reionize the universe \citep{ouchi:09,mitra:13,robertson:15}.   Ideally we would track SELGs from $z\sim0$ to the first galaxies at $z>8$.  However, current near-infrared instruments place a redshift limit of $z\sim4$ for identifying \OIII5007\AA\ emitters which are the focus of our study.

An effective method to identify galaxies with strong \OIII5007\AA\ emission (\ewrest$>200$\AA) at $z\gtrsim2$ is to first use deep multi-band photometry to select candidates and then confirm with near-IR spectroscopy.  The ELGs with the strongest \OIII5007\AA\ emission tend to be low-mass \citep[\logMstarMsun$<9.5$; e.g. ][]{maseda:13,maseda:14} systems, thus sensitive multi-wavelength imaging is needed to push down in stellar mass to select candidates.  Precise photometric redshifts at $z>1$ also are essential for identifying strong emission line features in the Spectral Energy Distributions (SEDs), and this requires medium-band near-IR imaging.  Lastly, only with deep near-IR spectroscopy can the extreme \OIII5007\AA\ nature of these systems be confirmed. 

With the advent of deep near-IR imaging surveys and sensitive near-IR spectrographs, we are now able to identify these strong \OIII5007\AA\ emitting galaxies at $z\sim3-4$.  Our method is similar to studies that couple near-IR imaging and near-IR spectroscopy to identify galaxies with strong equivalent widths at $z\sim1-2$, e.g. {\tt 3D-HST} \citep{maseda:13,maseda:14,maseda:18}.  First we use the \zfourge\ survey that measures precise photometric redshifts to $\sim70,000$ objects by combining deep imaging with medium-band near-IR filters \fsjhk\ and public multi-wavelength observations \citep[redshift  uncertainties of \sigmaz$\sim1.6$\%;][]{straatman:16}.  At $z\sim3$, the \zfourge\ survey is 80\% mass-complete to \logMstarMsun$\sim9.5$ and measures star formation rates down to $\sim5$~\Msunyr\ \citep{tomczak:16}.  

With photometry spanning observed UV to mid/far-IR, we then construct composite SEDs that are defined by the underlying galaxy populations \citep{kriek:11,forrest:16}.  In our analysis of \zfourge\ galaxies at \zphotrange, we discovered a population of $\sim80$ galaxies with blended rest-frame \HbOIII\ equivalent widths in excess of $\sim800$\AA\ \citep{forrest:17}.  In comparison, there are only $\sim14$ galaxies with such extreme \HbOIII\ at $1<z<3$ \citep{forrest:18}.  The rapid increase in the number density of the extreme \HbOIII\ emitting galaxies from $z\sim2$ to $z\sim3.5$ suggests potentially explosive growth at higher redshift \citep[see also][]{vanderwel:11}.

To spectroscopically confirm the \HbOIII\ emitting galaxies identified in \zfourge, we introduce our Multi-Object Spectroscopic Emission Line (\mosel\footnote{Mosel is also one of the 13 official German wine regions (Weinbaugebiete) and known for Riesling and Pinot Noir.}) survey.  In this paper, we focus on Emission Line Galaxies (ELGs) at \zphotrange\ to measure their redshifts and rest-frame \OIII5007\AA\ equivalent widths and line-widths.  We combine our spectroscopic measurements with physical properties obtained from deep multi-band imaging to infer gas fractions and virial masses, and test disk formation models.

By identifying Strong ELGs up to $z\sim4$ \citep{forrest:17,forrest:18}, \mosel\ complements existing emission line searches conducted with the {\it Hubble Space Telescope}.   Due to the wavelength ranges of the WFC3 and ACS grisms, blind spectroscopic surveys such as {\tt 3D-HST} \citep{momcheva:16}, {\tt WISP} \citep{atek:11}, and {\tt PEARS} \citep{straughn:08} are limited to SELGs at $z\lesssim2.3$.  Our medium-band NIR imaging from \zfourge\ combined with public legacy datasets enables us to reach comparable stellar masses as the blind spectroscopic surveys \citep[\logMstarMsun$\sim8.5$ at $z\sim1$;][]{straatman:16}.  At $z\sim3-4$ , we also span comparable ranges in rest-frame equivalent width ($\gtrsim200$\AA) and spectral line flux ($\sim1-2\times10^{-17}$\ergu) as the lower redshift studies. 

In our analysis, we use AB magnitudes and the galaxy parameters measured by \citet{forrest:17,forrest:18} for the \zfourge\ data-set.  FAST \citep{kriek:09a} is used to fit the SEDs assuming a Chabrier Initial Mass Function and an SED library with 1/5 solar metallicity and emission lines \citep[see][]{salmon:15,forrest:18}.  We assume $\Omega_{\rm m}=0.3$, $\Omega_{\Lambda}$=0.7,  $H_0=70$~\kms~Mpc$^{-1}$, and a flat Universe; the corresponding angular scale at $z=3.0$ is 7.7 kpc per arcsec.  

\section{Data \& Methods}

\subsection{Selecting Emission Line Galaxies }

The following summarizes the \zfourge\ observations we used to measure
photometric redshifts and galaxy properties as well as to generate the
composite SEDs.  For complete descriptions of the data products used
here, we refer the reader to the \zfourge\ survey paper by
\citet{straatman:16} and analysis of star formation rates by
\citet{tomczak:16}.

\subsubsection{\zfourge\ Imaging Catalogs}

We use the deep near-IR imaging from the FourStar Galaxy Evolution
survey \citep[\zfourge;][]{straatman:16} obtained with the FourStar
imager \citep{persson:13} on the Magellan Telescope of three legacy
fields: CDFS \citep{giacconi:02}, COSMOS \citep{scoville:07}, and UDS
\citep{lawrence:07}.  \zfourge\ divides the \textit{J}-band filter
into $J_1$, $J_2$, and $J_3$ and the \textit{H}-band filter into $H_s$
and $H_l$; \zfourge\ also obtains deep \Ks\ imaging that is used as
the detection image.  In combination with existing multi-wavelength
observations, \zfourge\ provides high precision photometric redshifts
with \sigmaz$=0.016$ \citep{straatman:16} for over 70,000
objects; the redshift precision is confirmed by the \zfire\
spectroscopic survey \citep{nanayakkara:16}.

We incorporate HST imaging from CANDELS \citep{grogin:11,koekemoer:11}
spanning 0.3$\mu$m to 1.6$\mu$m to measure photometry and galaxy
properties.  We also use \spitzer/IRAC (all four channels) and MIPS
data (\mipsmu) for the CDFS, COSMOS, and UDS fields (GOODS-S: PI
Dickinson, COSMOS: PI Scoville, UDS: PI Dunlop), and 100 and 160$\mu$m
for CDFS.  For CDFS only, we include public \textit{Herschel}/PACS
data \citep{elbaz:11}.  Total star formation rates are calculated by
combining the UV and IR contributions; see \citet[\S2]{tomczak:16} for
a detailed description. 

\subsubsection{Candidate Emission Line Galaxies at \zphotrange}\label{sec:types}

\zfourge\ is particularly sensitive to emission line galaxies at \zphotrange\ because \HbOIII\ emission falls in the deep \Ks\ imaging.  To identify galaxies with the strongest emission lines, we use the composite SEDs generated by \citet{forrest:17}.  From testing multiple fitting methods, \citet[\S4.5]{forrest:18} show that restframe equivalent widths down to $\sim20$\AA\ can be recovered from the composite SEDs; in the case of \HbOIII, the detection threshold applies to the blended \ewrest.  


To summarize, we iteratively select the \textit{primary galaxies} with the largest number of \textit{analog galaxies} based on the similarity of 22 rest-frame UVOIR colors \citep[$b<0.05$, from][]{kriek:11,forrest:16}  to collectively form separate \textit{composite groups}.  Observed photometry from analog galaxies in each composite group are then de-redshifted, normalized to a common flux scale, and combined to build a composite SED, essentially a low resolution ($R\sim40$) spectrum.

We focus on the two composite SEDs from \citet{forrest:17,forrest:18} with the steepest UV slope and strongest blended \HbOIII\ emission.  We adopt an admittedly arbitrary definition and refer to these emission line galaxies as {\it Strong} (SELG).  In our analysis, we refer to the following types of galaxies:

\begin{itemize}

\item Star-Forming Galaxy (SFG): composite SEDs with rest-frame \HbOIII\ equivalent widths of $<230$\AA\

\item Strong Emission Line Galaxy (SELG): the combined sample of 278 galaxies in \eelg\ and  \selg. 

\item \eelg:  the 60 galaxies grouped in the the composite SED with rest-frame \HbOIII\ equivalent width of $>800$\AA\

\item \selg:  the 218 galaxies grouped in the the composite SED with rest-frame\HbOIII\ equivalent width of $230-800$\AA 

\end{itemize}

Across the three \zfourge\ fields, we identify a total of 278 candidate Strong ELGs from the \zfourge\ composite SEDs, the majority of which are in CDF-S \citep{forrest:17}.  Except where noted, we use the combined sample of SELGs=(\eelg+\selg) .  We exclude Active Galactic Nuclei (AGN) identified by \citet{cowley:16} using multi-wavelength diagnostics; we discuss this in more detail in \S\ref{sec:sfr_agn}.

\subsection{Keck/MOSFIRE Spectroscopy}

\subsubsection{Observations}

We used MOSFIRE \citep{mclean:12} on Keck I (project code Z245, PI Kewley) on 12 and 13 February 2017.  We observed 5 masks in COSMOS and 1 mask in CDFS.   To measure \HbOIII\ at $z\sim3$, we used the K-band with wavelength range of $1.93-2.38\mu$m and spectral dispersion of 2.17~\AA/pixel.  We used an AB dither pattern with $1.5$ arcsecond nod and integrated each mask for a total on-sky time of 96 minutes (110 minutes clock time); seeing ranged from $0.7-1$ arcsec.

In the six MOSFIRE masks, we targeted a total of 105 galaxies at $0.9<$\zphot$<4.8$ where the highest priority targets were the 38 Strong ELGs candidates at \zphotrange.  The remaining targets (67) primarily were star-forming galaxies at $2<$\zphot$<4$.  Each mask included a flux monitor star to anchor the spectro-photometric calibrations.  We follow the same reduction pipeline as in our \zfire\ survey \citep{tran:15,nanayakkara:16,tran:17} and estimate a $3\sigma$ line-flux limit in the MOSFIRE K-band of $\sim3\times10^{-18}$~erg~s$^{-1}$~cm$^{-2}$.  Note that the angular sizes of the galaxies are comparable to or smaller than the slit-width of $0.7''$ (Fig.~\ref{fig:Mstar_size}), $i.e.$ there should be no significant systematic error such as slit-loss due to the spectro-photometric calibration.

\subsubsection{Spectroscopic Redshifts}\label{sec:zspec}

\begin{figure}
\plotone{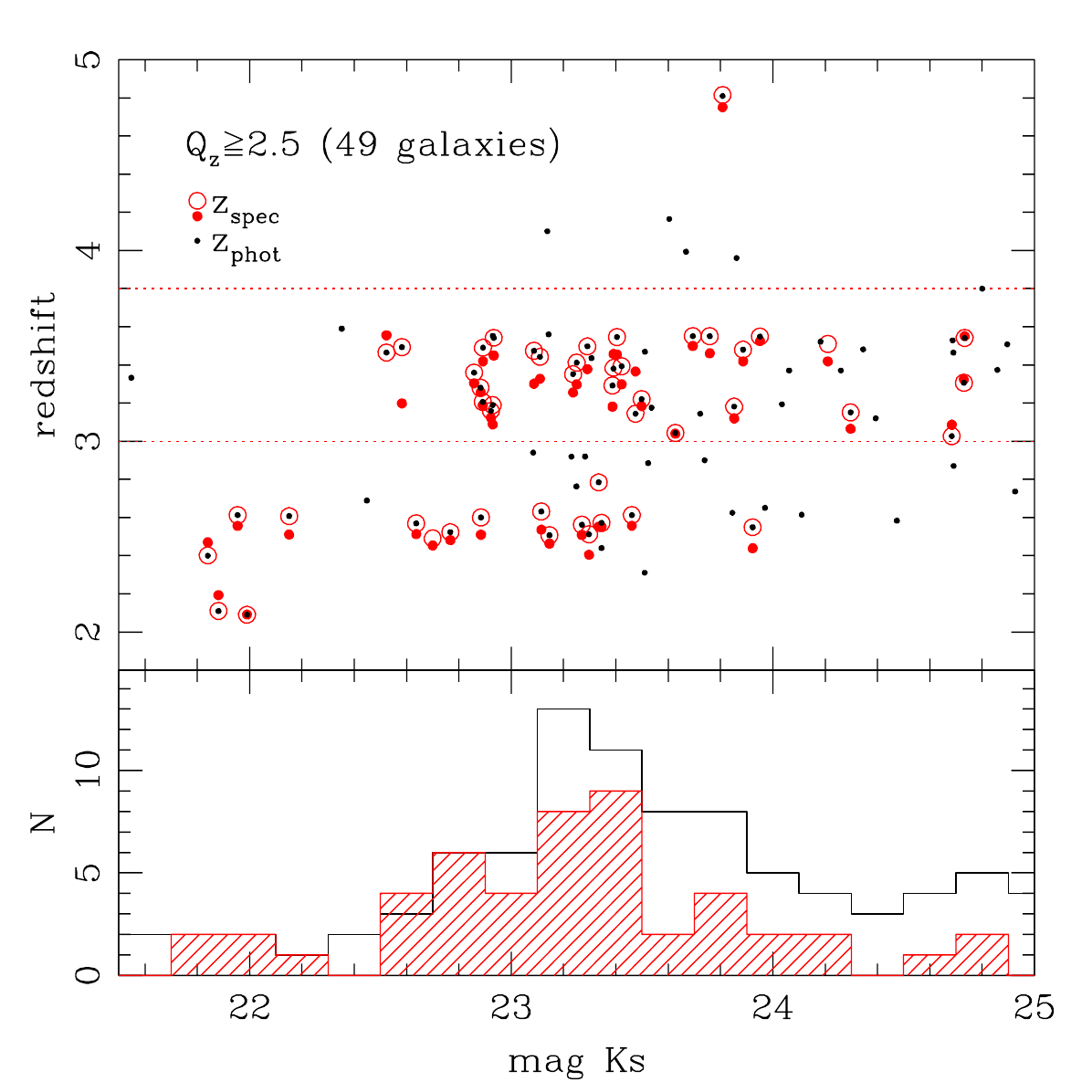} 
\caption{We compare the photometric redshifts (top) and \Ks\
  magnitudes (bottom) of the 95 targeted galaxies (small filled black circles \& open black
  histograms) to the \nzspec\ galaxies with spectroscopic redshifts
  (vertical pairs of large filled+open red circles \& hashed red histograms).  The two samples are likely
  drawn from the same parent population: KS tests comparing the
  \zphot\ and \Ks\ magnitudes of the \nzspec\ galaxies with \zspec\ to
  that of the targeted sample measure probabilities of 14\% and 6\%
  respectively.  The median \zspec\ of these \nzspec\ galaxies is only
  $\sim1$\% lower than their median \zphot\ (3.19 vs. 3.22). Our
  analysis focuses on the \nzspecOIII\ spectroscopically confirmed
  galaxies at \hizspecrange\ where \HbOIII\ fall in the MOSFIRE K-band
  (top panel, horizontal dashed lines).}
\label{fig:zhist}
\end{figure}

Of the 105 galaxies targeted with MOSFIRE, we measure spectroscopic redshifts for \nzspec\ (\zspecrange; Fig.~\ref{fig:zhist}).  Considering only the 89 targeted galaxies with photometric redshifts in the same range, i.e. galaxies where \OIII\ and \Halpha\ fall in the K-band, our success rate is $\sim53$\%.  The \nzspec\ galaxies with spectroscopic redshifts have \Ks\ magnitudes brighter than $25$ and quality flag of \Qz$\geq2.5$.  In our analysis, \Qz$\geq2.5$ means that the spectral line emission matches the \zfourge\ photometric redshift or there are two spectral lines with the same redshift \citep[for all definitions of \Qz, see][]{nanayakkara:16}.

On average, \zphot\ is $\sim0.054$ higher than \zspec\ (Fig.~\ref{fig:zhist}) and the corresponding uncertainty is \sigmaz=$[\Delta z/(1+z)]=0.0135$.  The largest outliers have $\Delta z\sim0.3$ and \sigmaz$\sim0.07$.  A two-sample Kolmogrov-Smirnov test shows the probability that the spectroscopically confirmed sample and the targeted sample are drawn from the same parent \zphot\ distribution to be 3.6\%, $i.e.$ the two distributions are different at the $2\sigma$ level.  The spectroscopically confirmed sample also is $\sim0.25$ magnitudes brighter with a K-S probability of being drawn from the same parent \Ks\ distribution as the targeted sample of 15\%.

Of the 13 targeted \eelg\ that were grouped in the composite SED with the highest \HbOIII\ emission (\ewrest$>800$\AA), two were lost due to mechanical problems with configuring the mask and two had no measured redshift.  The median redshift of the \nzEELG\ confirmed ELGs is \zspec=3.189 compared to their median \zphot=3.207.  The corresponding uncertainty of \sigmaz=0.42\% is even lower than that of our \zfire\ survey which targeted a broader selection of galaxies at $z\sim2$ \citep{tran:15,nanayakkara:16}.  Note also that the largest outliers are $\Delta z\sim0.1$ which is a factor of three smaller compared to the SFGs ($\Delta z\sim0.3$).  In our analysis, we focus on the 8 ELGs at \hizspecrange\ and exclude the ELG at \zspec$=2.549$ (ID 4791).

Of the 25 targeted \selg\ that were grouped in the composite SED with the second highest \HbOIII\ emission ($230<$\ewrest$<800$\AA), \nzSELG\ were spectroscopically confirmed with a median redshift of \zspec=3.327 compared to their median \zphot=3.41.  The corresponding uncertainty of \sigmaz=1.9\% for the \selg\ is larger than that of the \eelg\ and more typical of the \zfourge\ survey as a whole \citep{straatman:16,nanayakkara:16}.  All \nzSELG\ have redshifts of \hizspecrange.

Of the remaining 67 galaxies targeted with MOSFIRE, \nzSFG\ have spectroscopic redshifts (Fig.~\ref{fig:zhist}) with a median redshift of \zspec=2.551 compared to their median \zphot=2.612.  In our analysis, we focus on the 10 SFGs at \hizspecrange, i.e. we exclude the 17 galaxies at \zspec$<2.6$ and the one galaxy at \zspec$=4.815$.  We note that due to decreasing throughput of the near-IR arrays at $\lambda\gtrsim2.2\mu$m, the redshift cut-off is effectively $z\sim3.6$ except for objects with the strongest \OIII5007\AA\ emission.

\subsubsection{Measuring \OIII5007\AA\ Spectral Line Emission}\label{sec:linefitting}

\begin{figure}
\gridline{\fig{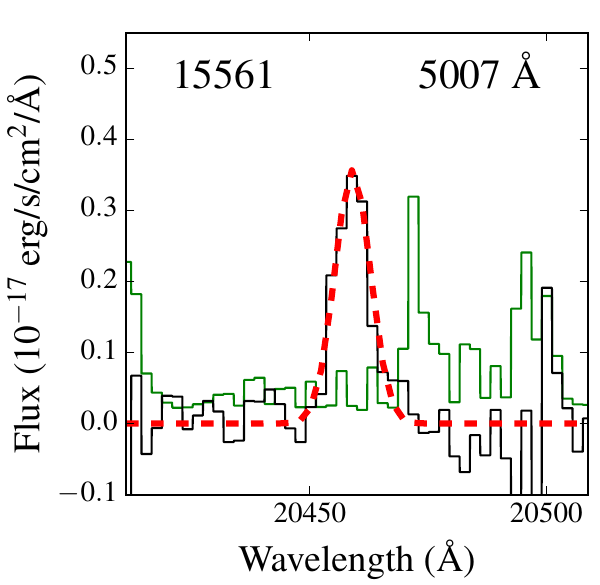}{0.2\textwidth}{CDFS}
          \fig{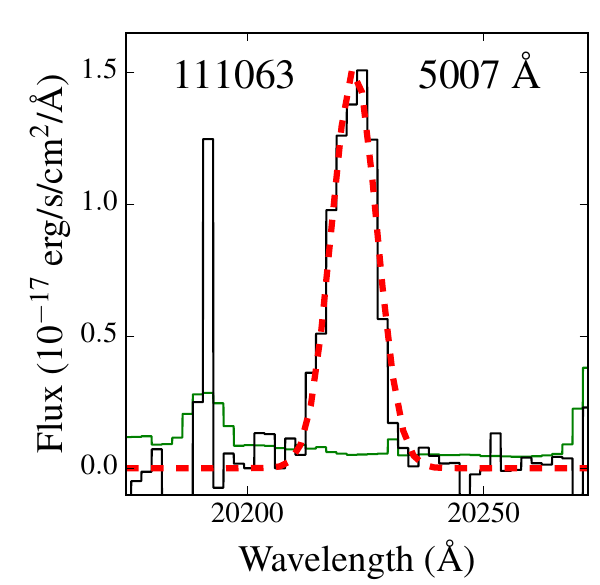}{0.2\textwidth}{COSMOS}
          \fig{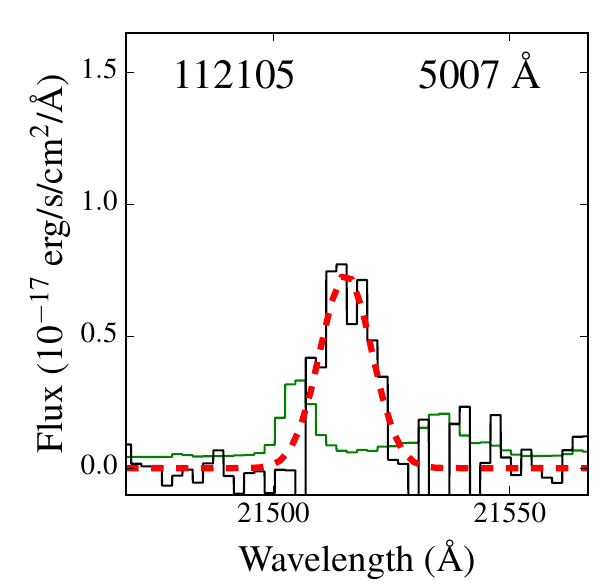}{0.2\textwidth}{COSMOS}
          \fig{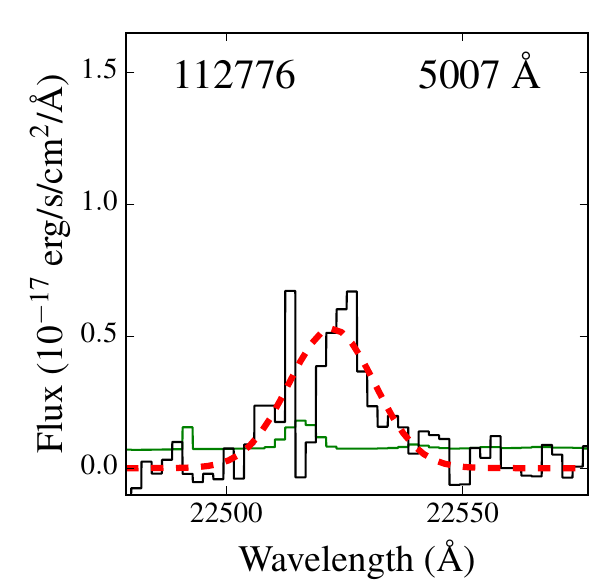}{0.2\textwidth}{COSMOS}
          \fig{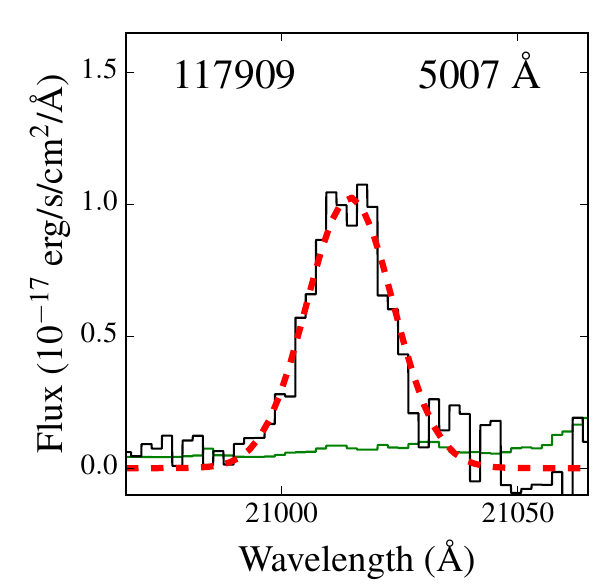}{0.2\textwidth}{COSMOS}}
\caption{Example of fits to the MOSFIRE spectra (\S\ref{sec:linefitting}) showing the observed 1D spectrum (black), the 1D error spectrum (green), and the 1D Gaussian fit (red dashed curve).}
\label{fig:linefit}
\end{figure}

Following \citet{alcorn:16,alcorn:18}, we first extract a 1D spectrum from an aperture defined by the $1\sigma$ Gaussian width of the \OIII5007\AA\ emission line along the spatial direction (Fig.~\ref{fig:linefit}).  To determine the [OIII]5007\AA\ line flux, we integrate the best-fit Gaussian centered on the line emission along the wavelength direction using the $3\sigma$ range; all line-fits are visually inspected for quality control.  We subtract in quadrature the instrumental broadening from the measured line-width and then convert to an integrated velocity dispersion (\intsigma) using the galaxy's measured redshift.  Errors in \intsigma\ are estimated by adding sky noise to the observed spectrum and refitting 1000 times.

\subsubsection{Determining \OIII5007\AA\ Equivalent Width}\label{sec:line_properties}

\begin{figure}
\plotone{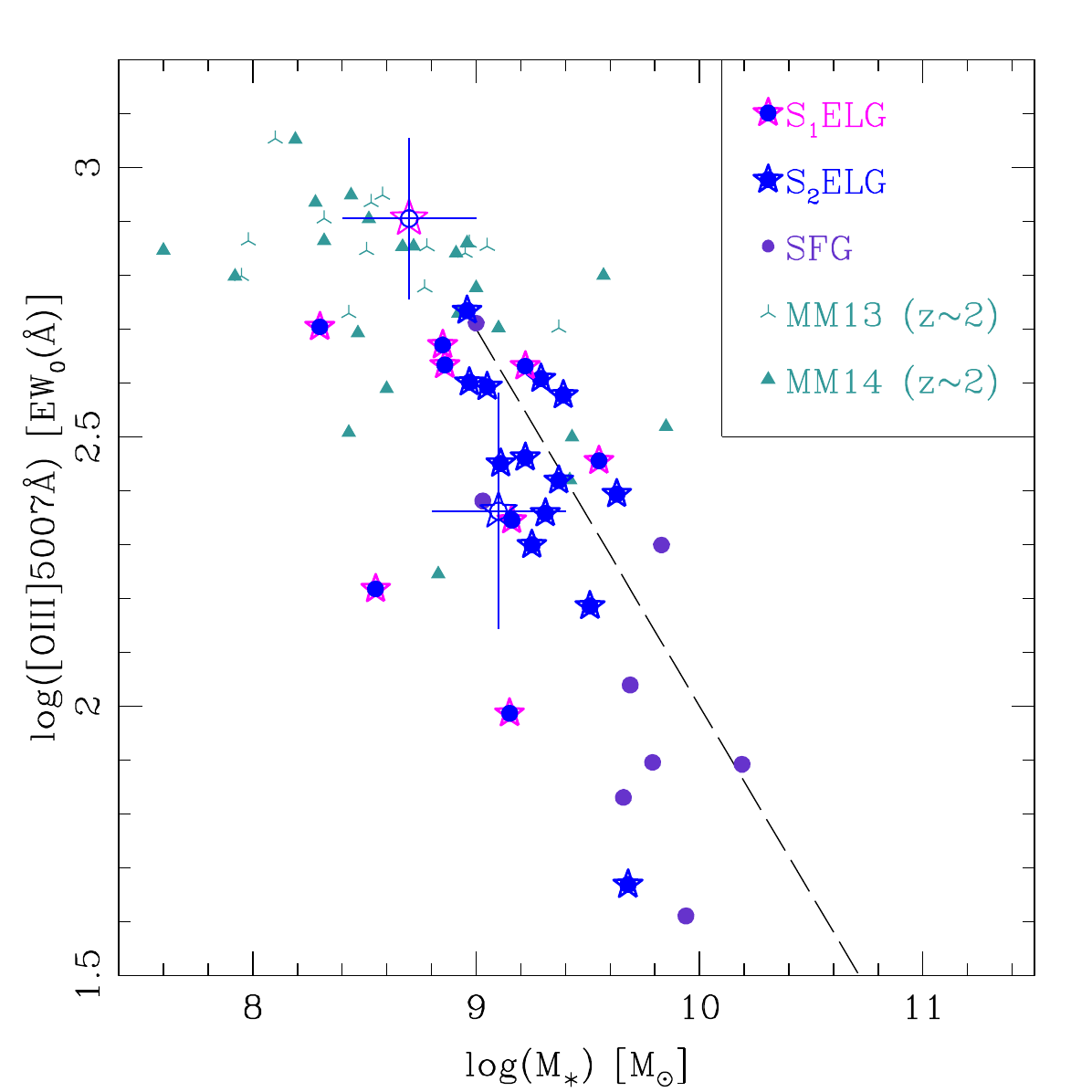} 
\caption{Spectroscopic rest-frame \OIII5007\AA\ equivalent widths for confirmed Emission Line Galaxies (ELGs) at \hizspecrange; typical uncertainties in stellar mass are $\sim2$ \citep{forrest:18}.  The symbols denote galaxy classifications as defined by the Composite SEDs, i.e. from photometry only, and  the large open stars show the rest-frame blended \HbOIII\ equivalent width {\it as measured from the two composite SEDs with the strongest emission} from \citet{forrest:17}.  Shown are: 1) \eelg\ with blended rest-frame \HbOIII\ \ewrest$>800$~\AA\ as measured from their Composite SED; 2) \selg\ with blended rest-frame \HbOIII\ \ewrest$\sim230-800$~\AA; and 3) more typical star-forming galaxies with blended rest-frame \HbOIII\ \ewrest$<230$~\AA.  For comparison, Strong ELGs at $1<z<3$ selected with {\tt 3D-HST} \citep{maseda:13,maseda:14} are shown in cyan; the spectral resolution and flux limit of the grism observations selects ELGs with the highest \OIII5007\AA\ EWs. The dashed line shows the average relationship between \OIII5007\AA\ \ewobs\ and stellar mass from MOSDEF for galaxies at $z\sim2$ \citep{reddy:18}. }
\label{fig:Mstar_eqw}
\end{figure}

To measure \OIII5007\AA\ equivalent widths, we require both line and continuum flux.  However, most of the ELGs are too faint to directly measure their continua from the MOSFIRE spectroscopy.  We use the method described in \citet{nanayakkara:17} that combines our spectro-photometrically calibrated \OIII5007\AA\ line fluxes with the  deep continuum photometry from \zfourge\ \citep{straatman:16}.  Because the \zfourge\ photometry provides a better measurement of the faint continuum relative to the spectroscopy, the primary source of uncertainty is thus due to systematic error of  the spectro-photometric calibration, and this uncertainty is of order $\sim10-20$\% for continuum-detected galaxies \citep{nanayakkara:16}.  Note that given the galaxy sizes are comparable to or smaller than the slit-width of (Fig.~\ref{fig:Mstar_size}), the systematic error due to the spectro-photometric calibration is not significant.

To determine the continuum for each ELG, we use the FAST fits \citep{kriek:09a} from \citet{forrest:17,forrest:18} that include a template library with strong emission lines.  As we show in both \citet{cohn:18} and \citet{forrest:18}, the stellar masses for low-mass galaxies (\logMstarMsun$\lesssim10$) can be overestimated by $\sim0.5-0.7$ dex if strong emission lines are not included in the SED modeling.  The emission lines are from \citet{salmon:15} who couple the photoionization code {\tt Cloudy} \citep{ferland:13} with BC03 simple stellar populations \citep{bruzual:03} as the ionizing source to generate nebular emission models.  

Because both star formation rate and stellar mass depend on the adopted stellar metallicity, SED fits are generated for solar ($Z=0.02$) and subsolar ($Z=0.004$) values.  The strong emission lines indicate the ELGs have gas metallicities lower than solar, thus we use the subsolar stellar metallicity fits \citep[$Z=0.004$; see also][]{cohn:18}.   However, we stress that the value adopted for the metallicity does not change the measured value for the continuum nor the measured \ewrest, only how we interpret the measurements.

We calculate the observed frame continuum on the blue and red side of the \HbOIII\ lines by using tophat filters (width of 150\AA) centered at 4675\AA\ and 5200\AA\ on the best-fit FAST SED.  We then divide the observed \OIII5007\AA\ line flux by the average observed continuum and the galaxy redshift.  

\begin{equation}
{\rm EW}_{\rm rest}(5007) = \frac{f_{\rm line}(5007)}{[(f_{\rm cont}(4675) + f_{\rm cont}(5200)]/2} \left(\frac{1}{1+z}\right)
\end{equation}

For a line flux of $3\times10^{-18}$~erg~s$^{-1}$~cm$^{-2}$ and continuum flux of $5\times10^{-20}$~\ergu\ (approximately \Ks\ magnitude of 24.0), the observed equivalent width is 60\AA; for a galaxy at \zspec=3.0, the corresponding rest-frame equivalent width is \ewrest$=15$\AA.  For comparison, the lowest values we measure for the spectral rest-frame equivalent widths using MOSFIRE are $\sim20$\AA\ (Table~2).  We note that \citet[\S4.5]{forrest:18} estimated rest-frame \ewrest\ down to $\sim20$\AA\ can be measured from the composite SEDs.

For reference, the {\tt 3D-HST} survey quotes a $3\sigma$ emission line flux limit for point sources of $1.5\times10^{-17}$~erg~s$^{-1}$~cm$^{-2}$ \citep{momcheva:16}.  Assuming the same continuum flux level, their limit corresponds to an observed equivalent width of 300\AA, i.e. approximately five times higher than \mosel.

\section{Results}

\begin{figure}
\plotone{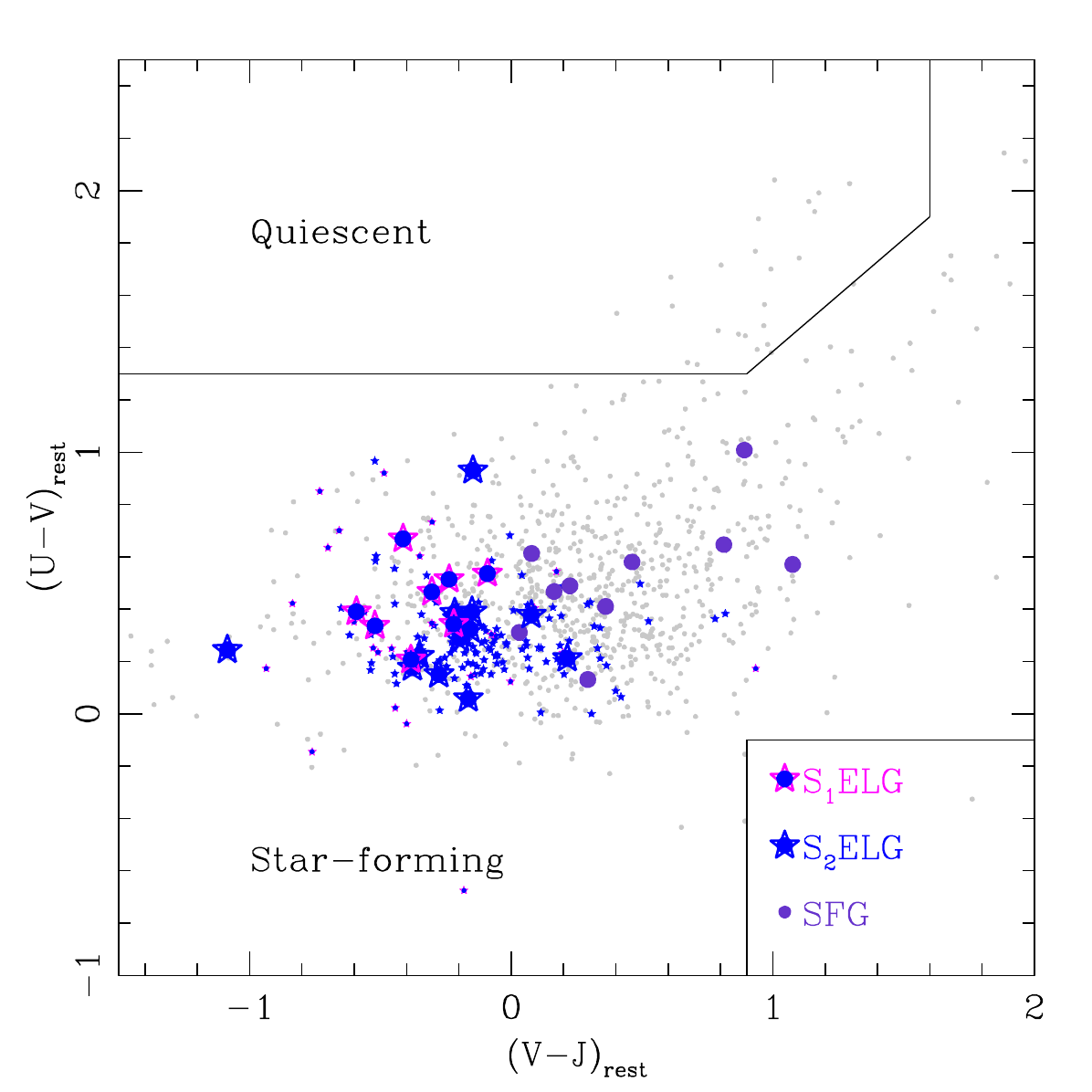} 
\caption{Rest-frame \uvj\ colors of \zfourge\ galaxies at \zphotrange\  identified using Composite SEDs, i.e. with photometry only \citep{forrest:17}.  The Strong ELGs are galaxies from the two composite SEDs with the strongest blended \HbOIII\ emission (\eelg\ and \selg; small stars).  Larger symbols show the \nzspecOIII\ spectroscopically confirmed  galaxies at \hizspecrange\ that include more typical star-forming galaxies (purple).  Typical uncertainties in rest-frame colors are not significant ($<0.1$), i.e. comparable to the symbol sizes.  Because of their strong \HbOIII\ emission, SELGs (stars) tend to have $(V-J)<0$ colors and are offset relative to the broader \zfourge\ population at \zphotrange\ (gray circles).  None of the \mosel\ galaxies are dusty as defined using the criterion from \citet{spitler:14} of \vmj$\geq1.2$.
}
\label{fig:uvj}
\end{figure}

In our analysis, we focus on the \nzspecOIII\ galaxies that are spectroscopically confirmed to be at \hizspecrange\ (Fig.~\ref{fig:zhist}).  We measure the \OIII5007\AA\ emission for these galaxies with our K-band spectroscopy (Fig.~\ref{fig:linefit}).  We combine our spectral measurements with deep photometry from \zfourge\ to measure continuum properties and use galaxy sizes from \citet{vanderwel:12}.  Although the \Luvir-based SF rates from \citet{tomczak:16} based on solar metallicity models are robust to significant flux from line emission, the stellar masses can be overestimated by as much as a factor of $\sim2$ \citep[e.g.][]{forrest:18,cohn:18}.  For these galaxies, we use stellar masses determined using updated FAST fits that include an SED template with strong emission lines and 1/5 solar metallicity \citep[$Z=0.004$;][]{forrest:18}.

\subsection{Strong \OIII5007\AA\ Emission}

With our MOSFIRE spectroscopy and deep multi-band imaging, we estimate rest-frame \OIII5007\AA\ equivalent widths using the hybrid method described in \S\ref{sec:line_properties}.  Our spectroscopically confirmed Strong ELGs span similar ranges with \OIII5007\AA\ \ewrest$\sim100-500$\AA\ (Fig.~\ref{fig:Mstar_eqw}).  These ranges are consistent with the large blended \HbOIII\ equivalent widths (\ewrest$\gtrsim 200$\AA) measured from their composite SEDs \citep{forrest:17,forrest:18}. 

The Strong ELGs at \hizspecrange\ show a trend of decreasing \OIII5007\AA\ equivalent width with increasing stellar mass that is also observed in SELGs at $z\sim2$ \citep[Fig.~\ref{fig:Mstar_eqw};][]{maseda:14}.  The significant overlap between \eelg\ and \selg\ indicates that the two are not distinctly different populations.  Note that the SELGs at $z\sim2$ include systems with \logMstarMsun$\sim8$ while our $z\sim3-4$ SELGs have \logMstarMsun$\gtrsim8.5$ due to sensitivity limits.  

The more typical star-forming galaxies (\HbOIII$\lesssim230$\AA) at \hizspecrange\ have larger stellar masses (\logMstarMsun$\gtrsim10$) and lower rest-frame \OIII5007\AA\ equivalent widths (\ewrest$\sim40-250$\AA; Fig.~\ref{fig:Mstar_eqw}).  This reflects the larger contribution of stellar continuum light, i.e. for two galaxies with the same \OIII5007\AA\ line-flux, the galaxy with the brighter continuum will have a lower equivalent width.  Our results confirm that selecting Strong ELGs from the \zfourge\ photometry is effective at identifying galaxies with the largest \OIII5007\AA\ equivalent widths.

For the \eelg, the \OIII5007\AA\ \ewrest\ values determined using the line fluxes obtained with MOSFIRE (see \S\ref{sec:line_properties}) tend to be lower than the \ewrest\ value estimated from the composite SED (Fig.~\ref{fig:Mstar_eqw}).  The offset is likely driven by how the continuum and emission lines are combined to generate the template used to fit the composite SEDs.  For example, underestimating the continuum will increase the inferred EW.  We refer the reader to \citet{forrest:17} who test three fitting methods on the composite SEDs of the strongest ELGs. 

\subsection{Rest-frame \uvj\ Colors}

With the \zfourge\ rest-frame wavelength coverage of $0.08-7\mu$m for each galaxy, we measure continuum properties including rest-frame \uvj\ colors from the individual SEDs \citep{tomczak:14,straatman:16}.  \zfourge\ galaxies at \zphotrange\ span the range in \uvj\ colors (Fig.~\ref{fig:uvj}) to include dusty and quiescent systems, but most lie in the star-forming region of the \uvj\ diagram \citep[see also][]{straatman:16}.  The ``typical'' star-forming galaxies in our spectroscopic sample have \vmj$\lesssim0.5$ that are values consistent with low amounts of reddening \citep[\Avsed$<0.5$;][]{forrest:16}.  None of the spectroscopically confirmed galaxies are dusty as defined using the criterion from \citet{spitler:14} of \vmj$\geq1.2$.

In contrast, the Strong ELGs are offset towards bluer \vmj\ colors (Fig.~\ref{fig:uvj}).  Their strong \HbOIII\ emission significantly boosts their $V$-band fluxes to produce rest-frame values of \vmj$<0$; this is particularly striking for the \eelg\ where virtually all have \vmj$<0$.  Such blue \uvj\ colors and non-detections in the far-IR indicate that these ELGs are essentially dust-free systems.  The relative distributions of the \eelg\ and \selg\  in the \uvj\ diagram suggests a continuum of phases where age and dust content increases from the Strong ELGs to the more typical star-forming galaxies, e.g. Lyman-Break Galaxies.

\subsection{Star-Formation Rate vs. Stellar Mass}

\begin{figure}
\plotone{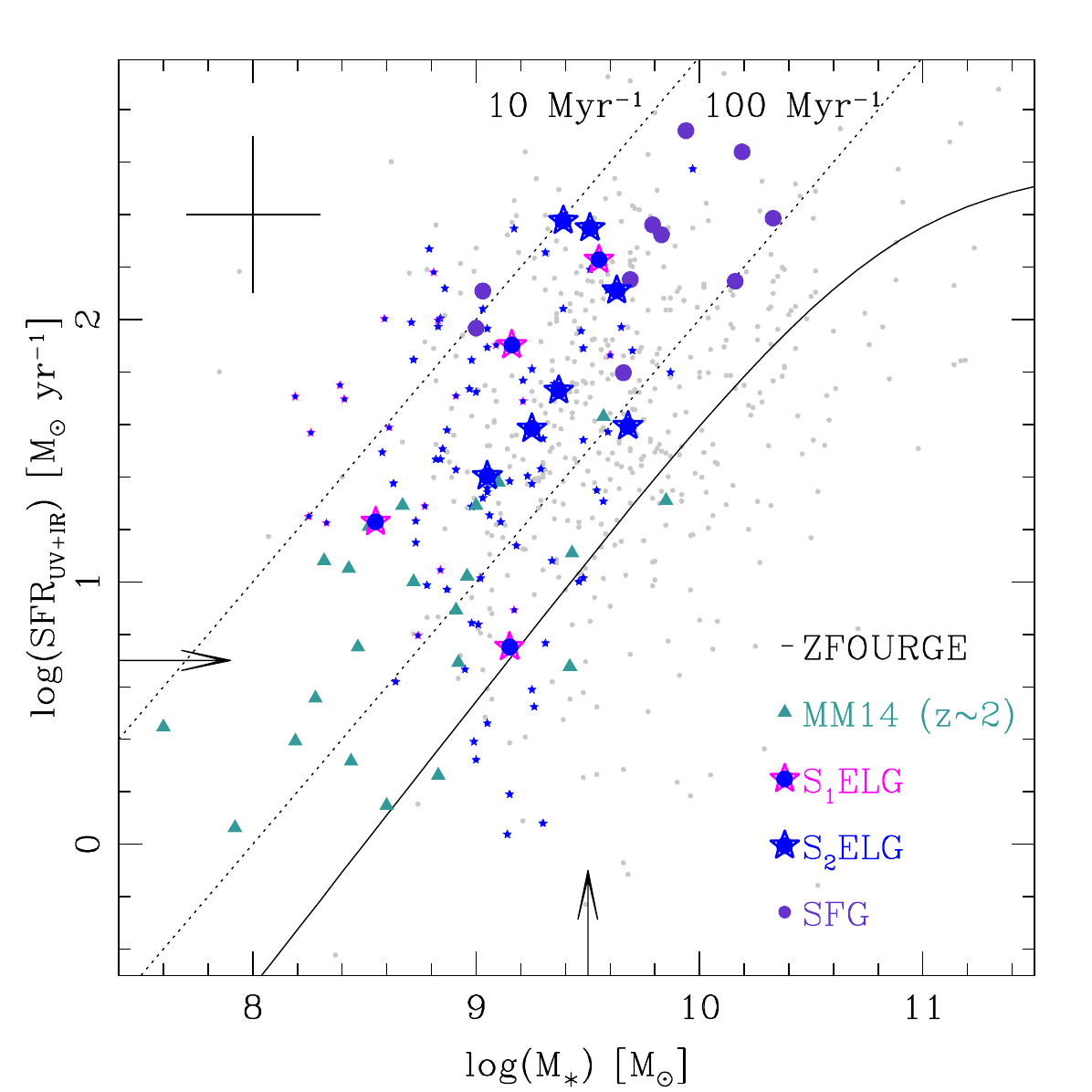} 
\caption{\HbOIII\ emitting galaxies tend to lie $\sim0.9$ dex above the star-forming main sequence (SFMS) at $z=3.5$ \citep[solid curve based on stacked star formation rates;][]{tomczak:16}; symbols are as in Fig.~\ref{fig:uvj} and illustrative errorbars correspond to a factor of two uncertainty in stellar mass and star formation rate.  Arrows denote the completeness limits at $z\sim3.5$ from \zfourge\ \citep{tomczak:16}.  Total star formation rates are based on \zfourge\ (UV+IR) fluxes, and here we plot only the galaxies that are individually detected in the IR.  The Strong \HbOIII-emitters with \zspec\ (large stars) have the  same distribution as their respective \zphot\ samples (small stars).  The diagonal dotted lines denote mass-doubling timescales of 10 and 100 Myr.  Strong ELGs at $1.4<z<2.3$ \citep[filled  triangles;][]{maseda:14} also tend to lie above the SFMS.}
\label{fig:Mstar_sfr}
\end{figure}

\begin{figure}
\plotone{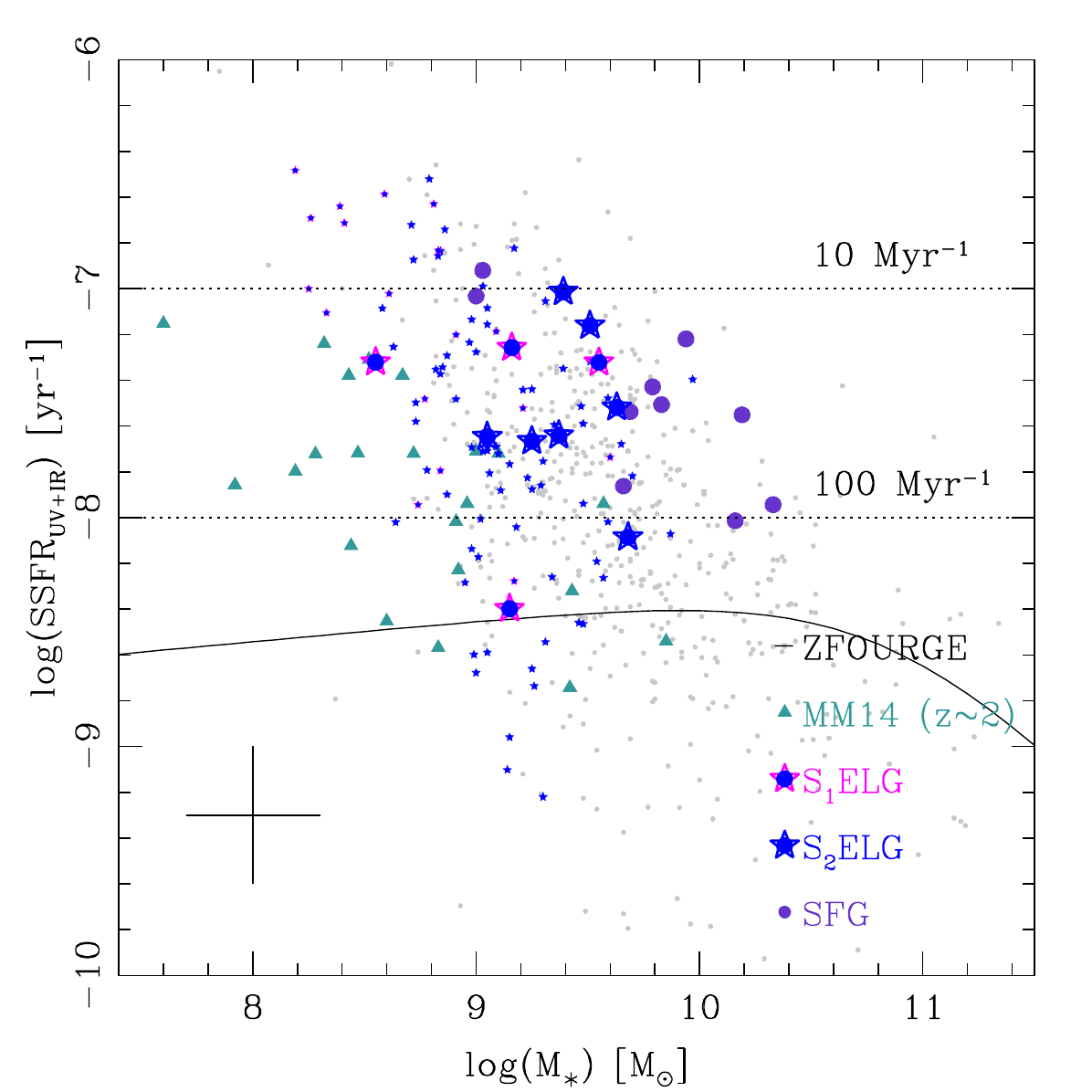} 
\caption{The starburst nature of the \HbOIII\ emitting galaxies is underlined by their high specific Star Formation Rates (SSFR) defined as their (UV+IR) SFRs divided by their stellar masses.  Symbols are as in Fig.~\ref{fig:Mstar_sfr} and included for comparison is the SSFR-\Mstar\ at $z=3.5$ from \citet{tomczak:16};  illustrative errorbars correspond to a factor of two uncertainty in stellar mass and specific star formation rate.  Most of the Strong ELGs (stars) have mass-doubling times of $<100$~Myr with several systems at $<10$~Myr (dotted horizontal lines).  Included for comparison are the ELGs at $1<z<3$  from \citet{maseda:14}  that all have mass-doubling times $>10$~Myr and tend to be low-mass (\logMstarMsun$<9$).  As SELGs grow in stellar mass and evolve into more typical SFGs, they move diagonally from the upper left to the bottom right.
}
\label{fig:Mstar_ssfr}
\end{figure}

The Strong ELGs tend to be lower mass systems [\logMstarMsun$\sim8.2-9.6$] compared to more typical star-forming galaxies (Fig.~\ref{fig:Mstar_eqw}).  At $z\sim3$, the $\log$(M$^{\star}$) for the stellar luminosity function from \zfourge\ is $\sim10.7$ \citep{tomczak:14}.  In comparison, the SELGs have stellar masses of only M$_{\star}\sim(0.003-0.08)$~M$^{\star}$.

Figure~\ref{fig:Mstar_sfr} shows the star formation rate to stellar mass (SFR-\Mstar) for the galaxies in our sample with measured (UV+IR) star formation rates from \zfourge\ \citep{tomczak:16}.  Although all of our galaxies have measured UV fluxes, many have negative IR fluxes due to the SED fitting method \citep[see \S2.5 of][]{tomczak:16} and thus 13 of the \nzspecOIII\ galaxies have negative (unphysical) total SFRs and are excluded from the SFR-\Mstar\ analysis.  Also, note that our $3\sigma$ line-flux limit in the MOSFIRE K-band is $\sim3\times10^{-18}$~erg~s$^{-1}$~cm$^{-2}$ \citep[see also][]{tran:17}.

Of the 18 ELGs with positive (UV+IR) SFRs, all lie above the relation between star formation and stellar mass commonly referred to as the Star-Forming Main Sequence (SFMS; Figs.~\ref{fig:Mstar_sfr} \& \ref{fig:Mstar_ssfr}); we confirm this is true even if we include UV only SFRs.  The ELGs tend to lie $\sim0.9$~dex above the SFMS at $z=3.5$ as measured by \citet{tomczak:16} from stacked SFRs based on (UV+IR) fluxes from \zfourge.  The overall distribution of the spectroscopically confirmed Strong ELGs mirrors that of the photometrically selected sample at this epoch, $i.e.$ most of the SELGs lie above the SFMS. 

With stellar mass-doubling time-scales of only $\sim10-100$~Myr, virtually all of the Strong ELGs are starbursts (Figs.~\ref{fig:Mstar_sfr} \& \ref{fig:Mstar_ssfr}).  Our results are consistent with \citet{amorin:17} and \citet{maseda:14} who find that Strong ELGs at $1<z<3$ have elevated SFRs for their given stellar mass.  At stellar masses of \logMstarMsun$\sim9-9.5$ where the two redshift samples overlap, our SELGs at \zphotrange\ have higher SFRs compared to the $z\sim2$ SELGs.  However, we note that for low mass galaxies (\logMstarMsun$\lesssim10$), the observed scatter in \Mstar-SFR increases with increasing redshift \citep{tomczak:16}.

A possible concern is that our (UV+IR) based SFRs are near or below the nominal IR detection limit at $z\sim3.5$ \citep{tomczak:16}.  However, the very lack of IR emission is consistent with SELGs having little to no dust.  We find additional support for the elevated SFRs and specific SFRs for the SELGs in \citet{cohn:18}:  using the SED fitting code \prospector\ \citep{conroy:09,leja:17}, \citet{cohn:18}  show that the Strong ELGs with \HbOIII\ \ewrest$\gtrsim800$\AA\ (\eelg) are dominated by the current starburst and have rising star formation rates. 

\subsection{Galaxy Size vs. Stellar Mass}

\begin{figure}
\plotone{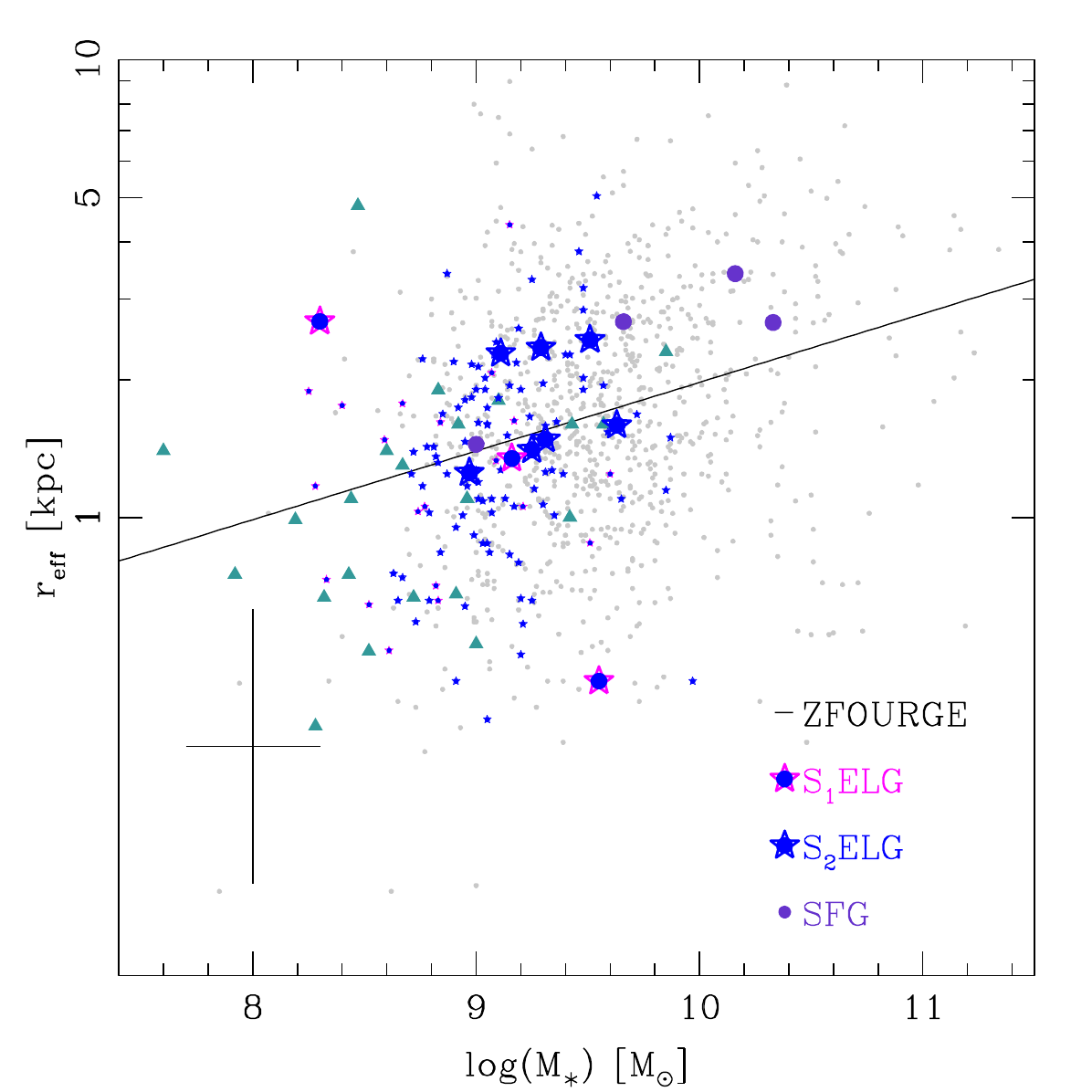} 
\caption{The \HbOIII\ emitters are consistent with the \zfourge\ galaxy size-stellar mass relation at $3<$\zphot$<3.75$ from \citet{allen:17} (solid line), but there is considerable scatter both in the ELGs and for all \zfourge\ galaxies at \zphotrange.  Here we use the effective radii measured by \citet{vanderwel:12}
  with WFC3/F160W imaging and the symbols are as in Fig.~\ref{fig:Mstar_sfr}.  The illustrative errorbars correspond to a factor of two uncertainty in stellar mass and galaxy size.  The strong ELGs at $z\sim2$ \citep[filled triangles;][]{maseda:14} are consistent with the same mass-size relation except at the lowest mass
(\logMstarMsun$<8.5$) where they tend to be more compact.}
\label{fig:Mstar_size}
\end{figure}

Our Emission Line Galaxies lie on the galaxy size-mass (\reff-\Mstar) relation measured by \citet{allen:17} using \zfourge\ galaxies at $3<$\zphot$<3.75$ (Fig.~\ref{fig:Mstar_size}).  Here we use the effective radii (galaxy size) measured by \citet{vanderwel:12} with the WFC3/F160W imaging and consider only galaxies with goodness of fit flag of 0.  These criteria further reduce our ELG sample to 13 galaxies.  We note that relaxing the goodness of fit flag to include all ELGs with measured \reff\ (28) increases the scatter in the galaxy size-mass relation but does not change the overall result.

Although they are virtually all starbursts (Figs.~\ref{fig:Mstar_sfr} \& \ref{fig:Mstar_ssfr}), our ELGs at \zphotrange\ follow the same \reff-\Mstar\ relation as the general population.  The Strong ELGs at $1<z<3$ \citep{maseda:14} are also consistent with the same size-mass relation except for the lowest mass systems (\logMstarMsun$<8.5$) which tend to lie below this relation, i.e. they are more compact at a given stellar mass.  Combining both redshift samples suggests that the Strong ELGs are an early phase in the continuum of stellar growth.

\subsection{Inferred Gas Fractions}

\begin{figure} 
\plotone{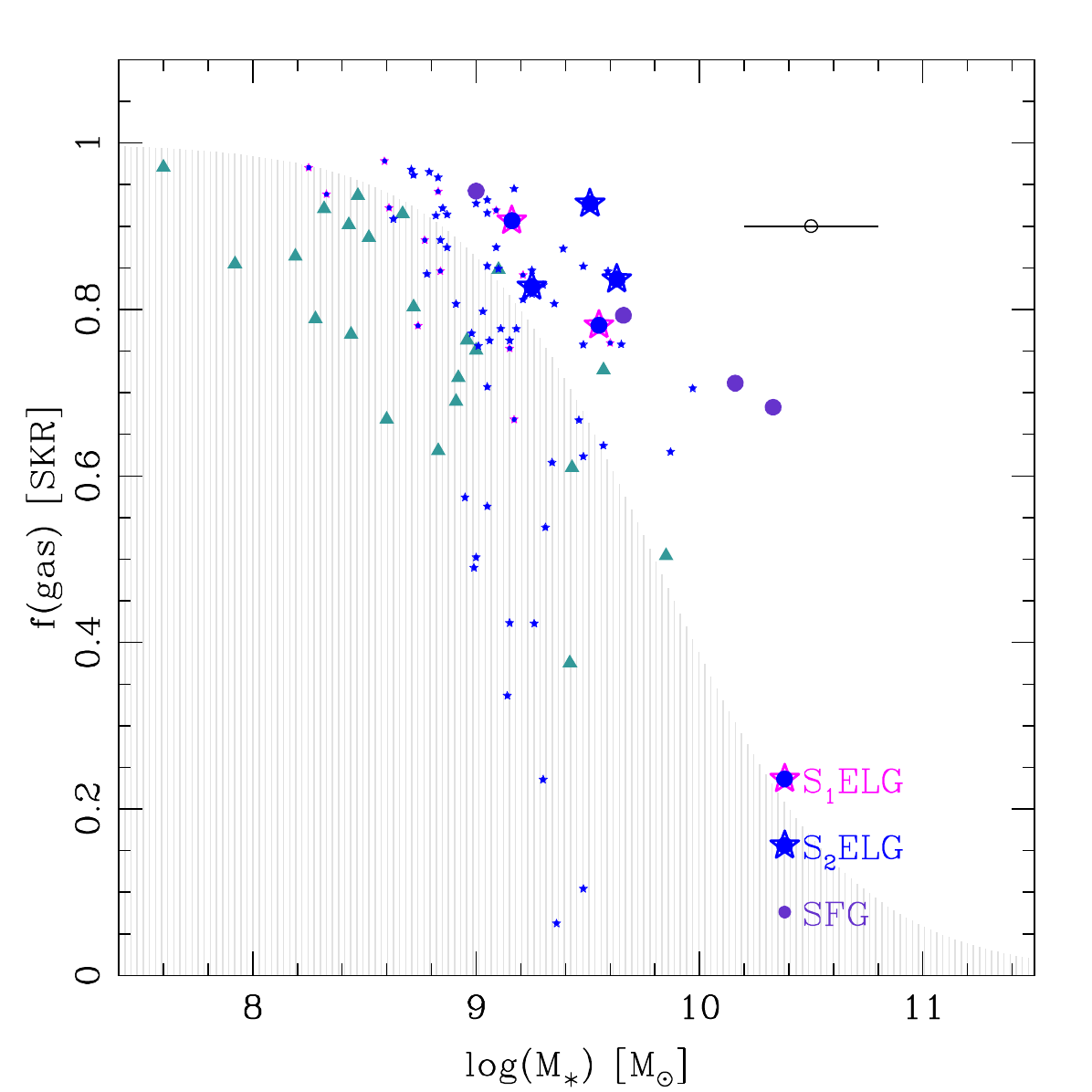} 
\caption{We infer gas masses by using the Schmidt-Kennicutt Relation \citep[SKR;][]{kennicutt:98} with effective sizes from vdW12 and total luminosities from \citet{tomczak:16}; symbols are as in Figs.~\ref{fig:uvj} \& \ref{fig:Mstar_sfr}.  The illustrative errorbars correspond to a factor of two uncertainty in stellar mass and we note that the gas mass is highly uncertain.  For reference, we show the exclusion region corresponding to a gas mass limit of \logMstarMsun$=9.8$ (light gray shaded region) which is below our nominal detection threshold at $z\gtrsim3$.  All of the spectroscopically confirmed ELGs in our study have inferred gas mass fractions of \fgas$>60$\%, but this is expected given the combination of their high specific star formation rates (see Fig.~\ref{fig:Mstar_ssfr}) and our detection limits at $z\sim3.5$.  The SELGs will move to the right as they increase in stellar mass and, unless their gas reservoirs are replenished, downwards.}  
\label{fig:Mstar_fgas}
\end{figure}

With UV+IR luminosities from \zfourge\ and \reff\ from the HST/F160W imaging \citep{vanderwel:12}, we use the Schmidt-Kennicutt Relation \citep[SKR;][Eq. 7]{schmidt:59,kennicutt:98} to estimate the gas surface density for individual galaxies:

\begin{equation}
\Sigma_{\rm SFR} = (2.5\pm0.7)\times10^{-4} 
\left( \frac{\Sigma_{\rm gas}}{1{\rm M}_{\odot}~{\rm pc}^{-2}}\right)^{1.4\pm0.15} 
{\rm M}_{\odot}~{\rm year}^{-1}~{\rm kpc}^{-2} 
\end{equation}

Assuming that half of the gas mass is within \reff, we use $\Sigma_{\rm SFR}$ to estimate the total gas mass:

\begin{equation}
\log({\rm M}_{\rm gas}) = 
\frac{5}{7} \log({\rm L}_{\rm UV+IR}) + \frac{2}{7} \log[\pi ({\rm r}_{\rm
eff})^2] + 1.52
\end{equation}

where \reff\ is measured in kpc, \Mgas\ in \Msun, and \Luvir\ in \Lsun; see also \citet{papovich:15}.  We use \reff\ defined by the stellar light; note that studies using CO \citep{tacconi:13} and \Halpha\ \citep{forster:11} find \reff\ from gas and stars are consistent.  Assuming an observational detection limit of \Luvir$=10^{11}$~\Lsun\ and typical galaxy size of \reff$=3.2$~kpc \citep{tran:17}, the corresponding gas mass limit is $\log$(\Mgas/\Msun)$=9.8$.  Gas fractions are defined to be \Mgas/(\Mgas+\Mstar).

All of our spectroscopically confirmed ELGS have inferred gas fractions of \fgas$>60$\% (Fig.~\ref{fig:Mstar_fgas}) which is not surprising given the ELGs' high specific star formation rates and our detection limits.  The high gas masses are consistent with high accretion rates that may be driving the star formation \citep{kacprzak:16}.   Our inferred gas fractions combined with measurements of Strong ELGs at $z\sim2$ \citep{maseda:14} suggests that \fgas$\gtrsim80$\% for ELGs with stellar masses of \logMstarMsun$<9$.   However, direct measurements of gas masses for the ELGs at $z>2$ with stellar masses of \logMstarMsun$<9$ will be difficult given current observational limitations.

\subsection{Kinematics}

\begin{figure}
\gridline{\fig{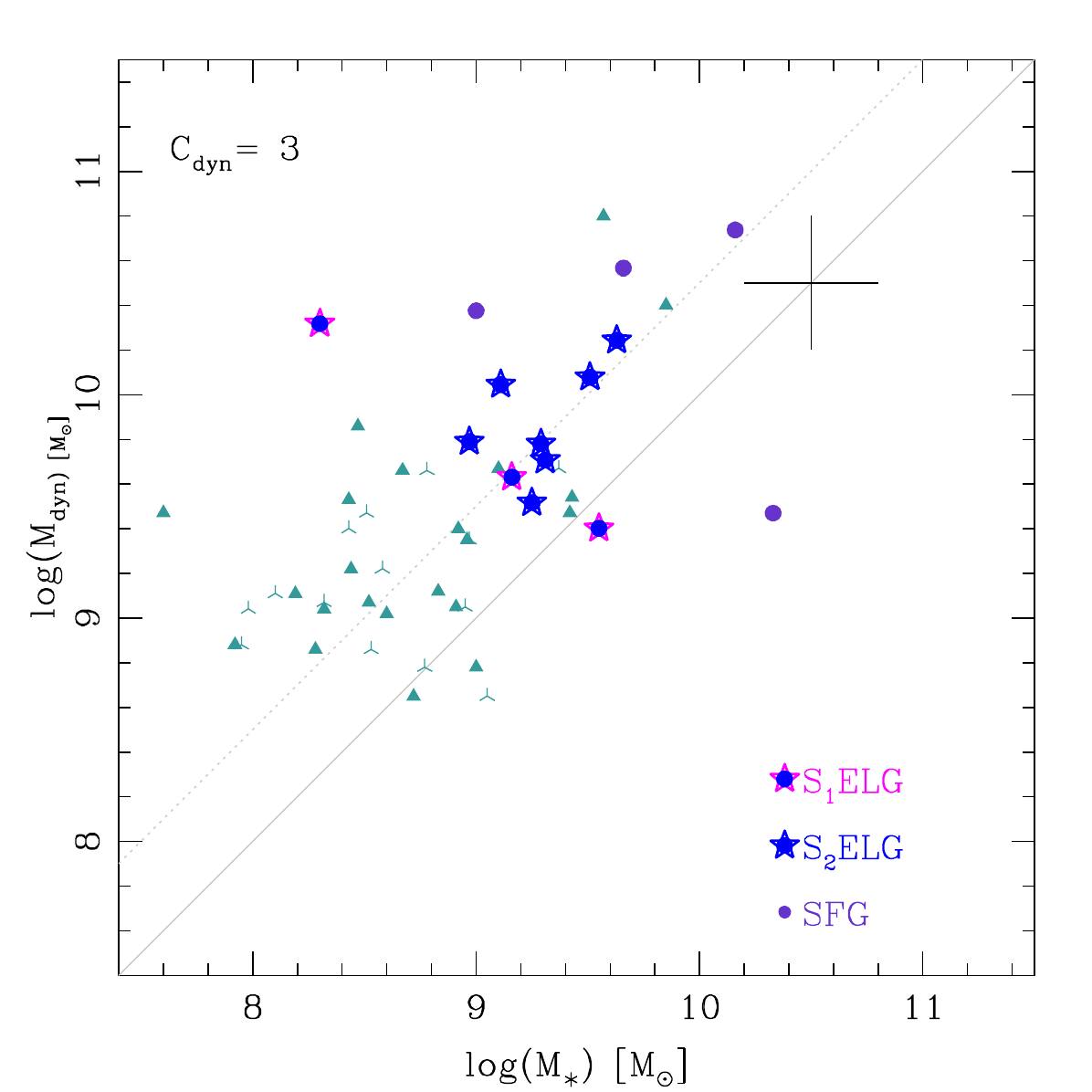}{0.5\textwidth}{}
          \fig{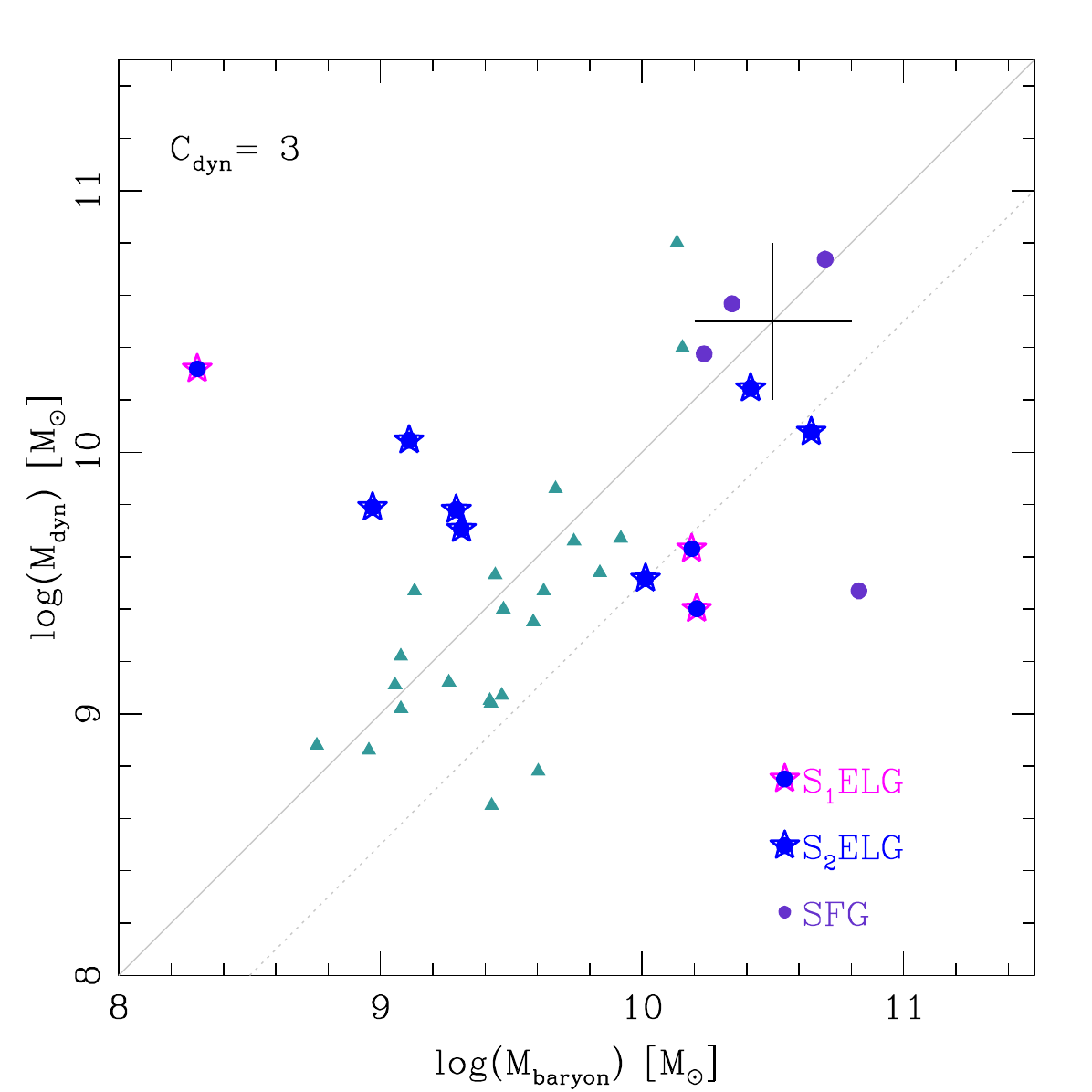}{0.5\textwidth}{}}
\caption{{\it Left:} The dynamical masses for the \OIII5007\AA\ emitting  galaxies at \hizspecrange\ are $\sim0.4$ dex larger than their stellar masses (symbols are as in Fig.~\ref{fig:Mstar_eqw}).    The illustrative errorbars correspond to a factor of two uncertainty in mass.  We estimate virial masses by combining the \OIII5007\AA\ line-widths with effective radii  measured by vdW12 \citep[see][]{alcorn:16,alcorn:18}.  The solid diagonal line  is parity and the dotted line is offset by 0.5 dex.  Our results are consistent with studies showing that ELGs at $z\sim2$ tend to have \Mvir$>$\Mstar \citep{maseda:13,maseda:14} and with their inferred gas mass fractions of \fgas$>60$\%.
{\it Right:} The same \mosel\ galaxies where we now compare the total baryonic mass (sum of the stellar and estimated gas mass) to their dynamical mass.  Including \fgas\ brings the \mosel\ galaxies closer to parity, but we hesitate to draw any stronger conclusions given the scatter and low number statistics.}
\label{fig:Mstar_Mvir}
\end{figure}

The integrated velocity dispersions (\intsigma) based on \OIII5007\AA\ line-widths is \intsigma$\sim40-150$~\kms\ for most of the ELGs with only one ELG having \intsigma$\sim200$~\kms\ (see Tables 1 \& 2).  Combining \intsigma\ and effective radii from vdW12, we follow \citet{alcorn:16} and estimate virial masses:

\begin{equation}
{\rm M}_{\rm dyn} (<{\rm R}_{e})= K_e \frac{\sigma^2_{\rm int} {\rm R}_e}{G}
\end{equation}

where for consistency with \citet{maseda:13,maseda:14}, we adopt the virial factor $K_e=3$ which is typically used for disk galaxies.

The dynamical masses for our ELGs at \hizspecrange\ are $\sim0.4$~dex larger than their stellar masses (Fig.~\ref{fig:Mstar_Mvir}); the handful of galaxies with \Mvir$<$\Mstar\ is consistent with scatter due to errors in the measurements.  The offset between virial and stellar mass is consistent with measurements of Strong ELGs at $z\sim2$ \citep{maseda:13,maseda:14} and continues the same trend to higher masses.  Within our limited sample at \hizspecrange, there is no obvious difference in the \Mstar-\Mvir\ relation for Strong ELGs compared to higher mass (\logMstarMsun$>9$) star-forming galaxies.  

When comparing the total baryonic mass (sum of the stellar and estimated gas mass) to dynamical mass, we find that the \mosel\ galaxies are closer to parity (Fig.~\ref{fig:Mstar_Mvir}, left).  However, given the scatter and low number statistics, we hesitate to draw any stronger conclusions regarding the ratio of dark to baryonic mass for the \mosel\ galaxies.

\subsection{Star Formation or Active Galactic Nuclei?}\label{sec:sfr_agn}

Our analysis assumes that the strong \OIII5007\AA\ emission is driven by star formation and not Active Galactic Nuclei (AGN).  We have used the \zfourge\ catalog by \citet{cowley:16} to remove AGN but recognize that at $z>3$, the multi-wavelength AGN diagnostics may not be reliable especially given the uneven coverage across these fields.  However, the \OIII5007\AA\ line-widths are consistent with star formation: most of the ELGs have \intsigma$\sim40-150$~\kms\ with only one ELG having \intsigma$\sim200$~\kms\ (see Tables 1 \& 2). 
Also, our recent results using \prospector\ to construct the star formation histories of the strong ELGs confirms that they are dominated by starbursts spanning the most recent $\sim50$ Myr \citep{cohn:18}.   Lastly, we note that  AGN contamination is rare in low-mass galaxies  \citep[e.g.,][]{ho:97,trump:15}.

Unlike the \zfourge\ composite SEDs where \Hbeta\ and \OIII5007\AA\ are blended \citep{forrest:17}, the MOSFIRE spectroscopy easily resolves these spectral features for individual ELGs.  Thus we also can identify potential AGN by combining the ratio of \OIII5007\AA\ to \Hbeta\ with stellar mass \citep{juneau:11}, although we note this method is contested at $z>1$ \citep{trump:13}.  The median \OIII5007/\Hbeta\ value for our sample of ELGs is $\sim5.8$ which is consistent with values reported by \citep{holden:16} for Lyman-Break Galaxies at $z\sim3$.  Following a similar line of analysis, \citet{maseda:14} also excluded AGN from their sample of strong ELGs at $1<z<2$.

\Hbeta\ is weaker than \OIII5007\AA\ and given our line-flux limit of $\sim3\times10^{-18}$erg~s$^{-1}$~cm$^{-2}$  ($3\sigma$), we can only place upper limits on the ratio of \OIII5007/\Hbeta\ for many of the ELGs.  A more careful treatment of the \Hbeta\ line-fluxes, e.g. by stacking the spectra, can be used to constrain ISM conditions.  Further analysis that includes \Hbeta, e.g. by combining \OIII5007/\Hbeta\ with stellar masses and star formation histories, will be presented in a future \mosel\ paper.  

\section{Discussion}

With deep multi-band photometry from \zfourge, we identified Emission Line Galaxies at $z>2.5$ that have blended rest-frame \HbOIII\ equivalent widths of $\gtrsim230$\AA\ \citep{forrest:17,forrest:18}.  We consider the combined sample of Strong ELGs grouped in the two composite SEDs with the largest \HbOIII\ equivalent widths (\ewrest$>230$\AA; see \S\ref{sec:types}).  The rarity of ELGs with \ewrest(\OIII5007)$\gtrsim200$\AA\ in the local Universe \citep[$\sim2$ ``green peas'' per square degree;][]{cardamone:09} raises the question of whether this Strong emission line phase is the exception or the norm at high redshifts.  In our \mosel\ survey, we build on recent studies to further explore how galaxies at $z\sim3.5$ with strong \OIII5007\AA\ emission fit into our current understanding of how star-forming galaxies grow by combining Keck/MOSFIRE K-band spectroscopy with our existing multi-band photometry from \zfourge.

\subsection{Strong \OIII5007\AA\ Emission May Be Common in Early Galaxy Formation}

We spectroscopically confirm \nzspecOIII\ galaxies at \hizspecrange\ with stellar masses of \logMstarMsun$\sim8.2-10.2$ and rest-frame \OIII5007\AA\ equivalent widths up to $\sim500$\AA\ (Figs. \ref{fig:zhist} \& \ref{fig:Mstar_eqw}).  The properties of the spectroscopically confirmed Strong ELGs mirror that of the larger photometrically selected sample (e.g. Figs. \ref{fig:uvj} \& \ref{fig:Mstar_sfr}).  Most of the SELGs have blue colors of \vmj$<0$ while the more typical star-forming galaxies have \vmj$\sim0-1$ (Fig.~\ref{fig:uvj}).  The overlapping ranges in their rest-frame \uvj\ colors suggest that the Strong ELGs transition into more massive star-forming galaxies, e.g. Lyman-Break Galaxies.

In the stellar mass range where we overlap with MOSDEF galaxies at $z\sim2$ \citep{reddy:18}, we find a similar relationship between \OIII5007\AA\ \ewrest\ and stellar mass (Fig.~\ref{fig:Mstar_eqw}).  \citet{reddy:18} suggest that the  increasing \OIII5007\AA\ \ewrest\ with decreasing stellar mass can be explained by either rapid enrichment of $\alpha$ elements or metallicities of $\lesssim0.2Z_{\odot}$ for galaxies with \logMstarMsun$\lesssim9$.  Both scenarios are consistent with our interpretation that the SELGs are young and have sub-solar metallicities. 

In combination with \citet{cohn:18} who show that SELGs at $z\sim3.5$ have low gas-phase metallicites ($Z_{\star}\lesssim0.02Z_{\odot}$) and higher specific star formation rates relative to SFGs (4.6 Gyr$^-1$ vs 1.1 Gyr$^-1$), our spectroscopic measurements support a scenario where strong \OIII5007\AA\ emission signals the earliest episodes of intense star formation \citep[see also][]{amorin:17}.   
As the SELGs grow in stellar mass, the growing amount of continuum light means that even during subsequent episodes of bursty star formation, the \OIII5007\AA\ equivalent widths will not be as large as during the first major burst of star formation.  With star formation rates of $\gtrsim3-250$~\Msun~yr$^{-1}$ (Fig.~\ref{fig:Mstar_sfr}) and mass-doubling times of $\sim10-100$~Myr (Figs.~\ref{fig:Mstar_sfr} \& \ref{fig:Mstar_ssfr}), the intense \OIII5007\AA\ emission phase is brief as these same galaxies quickly transition into more typical star-forming galaxies with \HbOIII\ \ewrest$\lesssim230$\AA.  

We find further support for a picture where strong \OIII5007\AA\ emission signals the earliest stages of stellar growth in galaxies by comparing relations between stellar mass (\Mstar), galaxy size (\reff), and virial mass (\Mvir).  Our SELGs follow the same general \Mstar-\reff\ relation as that of star-forming galaxies at $z\sim3$ (Fig.~\ref{fig:Mstar_size}), although we note the large scatter for all galaxies at this epoch.  The SELGs also continue the same trend between \Mvir-\Mstar\ as measured for SELGs at $z\sim2$ (Fig.~\ref{fig:Mstar_Mvir}).  The SELGs have virial masses that are larger by $\sim0.4$~dex relative to their stellar masses which is consistent with their inferred gas mass fractions of \fgas$>60$\% (Fig.~\ref{fig:Mstar_fgas}).

In a recent paper \citep{cohn:18}, we derived galaxy properties from the \zfourge\ photometry using the SED-fitting code \prospector\ \citep{leja:17} and the Flexible Stellar Population Synthesis package \citep[FSPS;][]{conroy:09}.  The \prospector\ code finds the best fit model and estimates uncertainties by sampling the posterior probability distributions of all the free parameters.   By calculating nonparametric star formation histories, \prospector\ can distinguish between rising, falling, and bursty star formation histories.

Using \prospector, \citet{cohn:18} show that ELGs with extreme emission (\eelg; \HbOIII\ \ewrest$\gtrsim800$\AA) are ``first burst'' systems and likely to have rising star formation rates. These same galaxies have low gas-phase metallicities of $Z_{\ast}\lesssim0.02Z_{\odot}$ and higher specific star formation rates compared to star-forming galaxies: $\sim4.6$~Gyr$^{-1}$ vs. $\sim1.1$~Gyr$^{-1}$.  \citet{cohn:18} inferred that many, if not most, star-forming galaxies at $z>2.5$ have Extreme \HbOIII\ emission-line phases early in their formation histories.  As these ``first burst'' systems continue to form stars and chemically enrich to evolve into more typical SFGs, they move diagonally from the upper left to the bottom right in Figs.~\ref{fig:Mstar_ssfr} \& \ref{fig:Mstar_fgas}.

\subsection{A Potential Source of Ionizing UV Photons}

A growing number of studies indicate that galaxies rather than AGN generated the UV photons needed to ionize the universe, but there are not enough massive galaxies at $z\gtrsim6$ to generate the required UV photons \citep{robertson:13,robertson:15}.  With several low-mass (\logMstarMsun$\lesssim9$) systems now identified at $z>3$ that have strong \ewrest(\OIII5007)$>300$\AA\ and escape fractions of \fesc$\gtrsim10$\% \citep{nakajima:16,debarros:16}, the most viable source of UV photons are these low-mass, star-bursting galaxies.  However, the stellar mass function at $z>8$ must be steeper than observed at lower redshifts for there to be enough of these dwarf galaxies to generate the required UV photons.

Another potential source of UV photons are the galaxies in our study with \ewrest(\OIII5007)$>200$\AA, e.g galaxies in a strong emission line phase.  The inferred gas fractions of \fgas$\gtrsim60$\% (Fig.~\ref{fig:Mstar_fgas}) and high specific star formation rates (Fig.~\ref{fig:Mstar_ssfr}) imply that the ELGs with the strongest \OIII5007\AA\ easily increase their stellar masses by factors $>2$ in less than $\sim100$~Myr, i.e. these Strong ELGs signal the earliest stages of stellar growth in galaxies \citep[see also][]{cohn:18}.

If the \OIII5007\AA\ emitters also have large \OIII5007/\OII3727\ ratios (O32$\gtrsim5$), studies indicate they may leak more Lyman-Continuum photons due to their harder ionizing spectrum \citep{izotov:16,nakajima:14,nakajima:16}.  \citet{tang:19} find that in the most intense line emitters at $z\sim2$,  the ionizing photon efficiency scales with \OIII5007\AA\ emission.   However, recent results by \citet{bassett:19} of galaxies at $z\sim3$ suggest that the correlation between large \OIII5007/\OII3727\ ratios and more Ly-C photons is weak at best, and \citet{naidu:18} constrain the average escape fractions for SELGs to be $8.5-16.7$\%.  

Only with spectroscopy can we measure fluxes of oxygen lines for individual galaxies to measure their ratios and determine what drives the strong \OIII5007\ emission, e.g. shocks or massive binary stars \citep{strom:17}.  
By obtaining at $z\sim3.5$ the ratio of \OIII5007\ to well-studied emission lines such as \OII3727, \Hbeta, and \Lya\ \citep[e.g.][]{tang:19,bassett:19}, we can better track how the ionizing photon efficiency evolves from the first galaxies to $z\sim0$.  We plan to measure \OII3727\AA\ emission for our ELGs to characterize their ionization conditions and constrain their production of Lyman-Continuum photons.

\section{Conclusions}

Our Multi-Object Spectroscopic Emission Line (\mosel) survey focuses on galaxies with strong \OIII5007\AA\ emission identified using deep broad-band photometry from the \zfourge\ survey \citep{forrest:17,forrest:18}.  We use Keck/MOSFIRE K-band spectroscopy and measure redshifts of \nzspec\ galaxies at \zspecrange.  Our spectroscopic success rate is $\sim53$\% and \zphot\ uncertainty is \sigmaz=$[\Delta z/(1+z)]=0.0135$ \citep[\S\ref{sec:zspec}, Fig.~\ref{fig:zhist}; see also][]{straatman:16,nanayakkara:16}.  

Of the \nzspec\ spectroscopically confirmed galaxies at \zspecrange, we measure \OIII5007\AA\ line fluxes for \nzspecOIII\ galaxies at \hizspecrange\ (Fig.~\ref{fig:linefit}).  By dividing the line-flux as measured with MOSFIRE by the continuum flux from \zfourge, we estimate rest-frame \OIII5007\AA\ equivalent widths of $\sim100-500$\AA\ where \ewrest\ increases with decreasing stellar mass (Fig.~\ref{fig:Mstar_eqw}).  Our analysis focuses on the Strong Emission Line Galaxies (SELGs) grouped in the two composite SEDs with the strongest \HbOIII\ emission (\ewrest$>230$\AA) from \citet{forrest:18}. 

We explore the properties of SELGs at $z\sim3.5$ to connect them to our current picture of star-forming galaxies.  The physical properties of the spectroscopically confirmed \OIII5007\AA\ Strong Emission Line Galaxies (SELGs) mirror that of the larger photometrically selected sample.  For example, the SELGs tend to have bluer colors of \vmj$<0$ compared to more typical star-forming galaxies with \vmj$\sim0-1$  (Fig.~\ref{fig:uvj}).

The Strong \HbOIII\ emitting galaxies in our study have stellar masses of \logMstarMsun$\sim8.2-9.6$ (Fig.~\ref{fig:Mstar_sfr}).  The same galaxies lie $\sim0.9$~dex above the star-forming main sequence at $z=3.5$ and have high specific star-formation rates with mass-doubling timescales of $\sim10-100$~Myr (Fig.~\ref{fig:Mstar_ssfr}).  The inferred gas fractions of \fgas$\gtrsim60$\% (Fig.~\ref{fig:Mstar_fgas}) can easily fuel a burst that increases stellar mass by $>2$.   In terms of stellar and virial mass, (UV+IR) star formation rate, and galaxy size, our \HbOIII\ emitting galaxies bridge relations measured for Strong ELGs at $1<z<3$ \citep[\logMstarMsun$\lesssim9$;][]{vanderwel:11,maseda:14}  to star-forming galaxies at $z\sim3.5$ (see Figs.~\ref{fig:Mstar_sfr}, \ref{fig:Mstar_ssfr}, \& \ref{fig:Mstar_size}).

Taken as a whole, our analysis suggests that strong \OIII5007\AA\ emission (\ewrest$\gtrsim200$) signals an early episode of intense star formation in low-mass (M$_{\star}<0.1$~M$^{\star}$) galaxies at $z\gtrsim3$.  The ELGs with the strongest \OIII5007\AA\ are a rapidly evolving population of galaxies both in number density and stellar growth \citep{forrest:17,cohn:18}.  The \OIII5007\AA\ ELGs are likely to evolve into more massive and older star-forming galaxies with stable disks and bulges, e.g. Lyman-Break Galaxies.  

In a recent paper \citep{cohn:18}, we estimated that many, if not most, star-forming galaxies at $z>3$ are strong \OIII5007\AA\ emitters early in their formation history.  If strong \OIII5007\AA\ emission is a common phase in early galaxy formation, this brief episode may generate a significant number of ionizing UV photons.  In a future paper, we will explore additional line diagnostics, e.g. the ratio of \OIII5007\AA\ to \Hbeta, to characterize ionization conditions and constrain the production of Lyman-Continuum photons in galaxies with the strongest \OIII5007\AA\ emission.

\acknowledgements

We are grateful to the Keck/MOSFIRE team with special thanks to M. Kassis, J. Lyke, G. Wirth, and L. Rizzi on the Keck support staff.  K. Tran thanks P. Oesch, B. Holden, and M. Maseda for helpful discussions, and B. Forrest thanks the Hagler Institute for Advanced Study at Texas A\&M for support.  We also thank the referee for a thoughtful and constructive report.  This work was supported by a NASA Keck PI Data Award administered by the NASA Exoplanet Science Institute. Data presented herein were obtained at the W. M. Keck Observatory from telescope time allocated to NASA through the agency's scientific partnership with the California Institute of Technology and the University of California. The Observatory was made possible by the generous financial support of the W. M. Keck Foundation.  K. Tran acknowledges that this material is based upon work supported by the National Science Foundation under Grant Number 1410728 and acknowledges the ARC Centre for Excellence in All-Sky Astrophysics in 3D (ASTRO 3D) for support in preparing the manuscript.  GGK acknowledges the support of the Australian Research Council through the DP170103470.  The authors wish to recognize and acknowledge the very significant cultural role and reverence that the summit of Mauna Kea has always had within the indigenous Hawaiian community. We are most fortunate to have the opportunity to conduct observations from this mountain.


\clearpage
\begin{deluxetable}{lrrrrrrrrrrrr}
\rotate
\tabletypesize{\scriptsize}
\tablecaption{\mosel\ Galaxy Properties\tablenotemark{a}}\label{tab:imaging}
\tablewidth{0pt}
\tablehead{
\colhead{Field} &
\colhead{\zfourge \tablenotemark{b}}          &
\colhead{RA \tablenotemark{b}} & \colhead {Dec \tablenotemark{b}} &
\colhead{\zspec} & \colhead{\zphot \tablenotemark{b}} &
\colhead{$K_{\rm obs}$ \tablenotemark{b}} &
\colhead{\umv \tablenotemark{b}} &
\colhead{\vmj \tablenotemark{b}} &
\colhead{\logMstarMsun \tablenotemark{c}}         &
\colhead{$\log$(SFR) \tablenotemark{c}} &
\colhead{\reff \tablenotemark{d}}
\\
\colhead{} &
\colhead{ID}          &
\colhead{J2000} & \colhead{J2000} &
\colhead{$\pm0.003$} & \colhead{$\pm0.0135$} &
\colhead{mag} &
\colhead{mag} &
\colhead{mag} &
\colhead{}         &
\colhead{\Msunyr} &
\colhead{kpc}
}
\startdata
COSMOS & 1877 & 150.170425 & 2.199359 & 3.1230 & 3.16 & 22.92 & 0.58 & 0.46 & 10.2 & 2.1 & 3.4 &  \\
COSMOS & 4214 & 150.177109 & 2.221284 & 3.4578 & 3.38 & 23.39 & 1.01 & 0.89 & 10.3 & 2.4 & 2.7 &  \\
COSMOS & 7239 & 150.202240 & 2.254340 & 3.1198 & 3.18 & 23.85 & 0.24 & -1.08 & 9.0 & $<1$ & 0.7 &  \\
COSMOS & 9884 & 150.072495 & 2.282637 & 3.2982 & 3.39 & 23.42 & 0.93 & -0.15 & 9.1 & 1.4 & 1.4 &  \\
COSMOS & 11063 & 150.146133 & 2.297038 & 3.0393 & 3.04 & 23.63 & 0.34 & -0.22 & 8.9 & $<1$ & 1.8 &  \\
COSMOS & 11284 & 150.136337 & 2.298915 & 3.3016 & 3.47 & 23.09 & 0.32 & -0.16 & 9.4 & 1.7 & 2.4 &  \\
COSMOS & 11544 & 150.147446 & 2.301592 & 3.3038 & 3.36 & 22.86 & 0.29 & -0.20 & 9.5 & 2.3 & 2.4 &  \\
COSMOS & 12000 & 150.070665 & 2.305136 & 3.2578 & 3.28 & 22.88 & 0.65 & 0.81 & 10.2 & 2.6 & \nodata &  \\
COSMOS & 12105 & 150.138840 & 2.306907 & 3.2976 & 3.41 & 23.25 & 0.54 & -0.09 & 9.2 & 1.9 & 1.3 &  \\
COSMOS & 12273 & 150.147030 & 2.309289 & 3.1809 & 3.29 & 23.39 & 0.31 & 0.03 & 9.8 & 2.4 & \nodata &  \\
COSMOS & 12776 & 150.121275 & 2.315327 & 3.4993 & 3.55 & 23.69 & 0.57 & 1.08 & 9.8 & 2.3 & \nodata &  \\
COSMOS & 12922 & 150.069214 & 2.315987 & 3.2556 & 3.35 & 23.24 & 0.47 & 0.16 & 9.7 & 1.8 & 2.7 &  \\
COSMOS & 14984 & 150.060333 & 2.338560 & 3.3777 & 3.50 & 23.29 & 0.18 & -0.38 & 9.3 & $<1$ & 1.5 &  \\
COSMOS & 15625 & 150.139206 & 2.345322 & 3.1841 & 3.22 & 23.50 & 0.36 & -0.20 & 9.2 & 1.6 & 1.4 &  \\
COSMOS & 15636 & 150.065445 & 2.345667 & 3.4188 & 3.51 & 24.21 & 0.13 & 0.29 & 9.0 & 2.0 & 1.4 &  \\
COSMOS & 16067 & 150.200134 & 2.349396 & 3.1885 & 3.21 & 22.89 & 0.47 & -0.30 & 9.2 & $<1$ & 1.4 &  \\
COSMOS & 16325 & 150.203055 & 2.352349 & 3.4538 & 3.55 & 23.40 & 0.49 & 0.22 & 9.7 & 2.2 & \nodata &  \\
COSMOS & 16513 & 150.066735 & 2.353337 & 3.4188 & 3.49 & 22.89 & 0.22 & -0.35 & 9.2 & $<1$ & \nodata &  \\
COSMOS & 16518 & 150.211500 & 2.354372 & 3.3653 & 3.14 & 23.47 & 0.41 & 0.36 & 9.9 & 2.7 & \nodata &  \\
COSMOS & 16984 & 150.083664 & 2.358806 & 3.3273 & 3.44 & 23.11 & 0.38 & 0.08 & 9.3 & $<1$ & 2.3 &  \\
COSMOS & 17008 & 150.168793 & 2.358994 & 3.4608 & 3.55 & 23.76 & 0.21 & 0.21 & 9.0 & $<1$ & 1.3 &  \\
COSMOS & 17423 & 150.115402 & 2.363473 & 3.5259 & 3.55 & 23.95 & 0.51 & -0.24 & 9.6 & 2.2 & 0.4 &  \\
COSMOS & 17909 & 150.094330 & 2.370171 & 3.1977 & 3.49 & 22.58 & 0.39 & -0.15 & 9.6 & 2.1 & 1.6 &  \\
COSMOS & 18022 & 150.079529 & 2.367794 & 3.4188 & 3.48 & 23.89 & 0.15 & -0.28 & 9.1 & $<1$ & 2.3 &  \\
COSMOS & 20001 & 150.214305 & 2.378608 & 3.4488 & 3.54 & 22.93 & 0.39 & -0.22 & 9.4 & 2.4 & \nodata &  \\
CDFS & 22136 & 53.152866 & -27.749243 & 3.0883 & 3.19 & 22.93 & 0.34 & -0.52 & 9.2 & 0.8 & 1.6 &  \\
CDFS & 15782 & 53.174133 & -27.800318 & 3.0651 & 3.15 & 24.30 & 0.61 & 0.08 & 9.0 & 2.1 & 0.2 &  \\
CDFS & 18053 & 53.195736 & -27.782713 & 3.3239 & 3.31 & 24.73 & 0.21 & -0.38 & 8.6 & 1.2 & 0.4 &  \\
CDFS & 17189 & 53.198280 & -27.789150 & 3.5506 & 3.54 & 24.73 & 0.39 & -0.59 & 8.3 & $<1$ & 2.7 &  \\
CDFS & 14864 & 53.204610 & -27.806757 & 3.5552 & 3.47 & 22.52 & 0.06 & -0.16 & 9.7 & 1.6 & \nodata &  \\
CDFS & 15561 & 53.219540 & -27.802586 & 3.0865 & 3.03 & 24.68 & 0.67 & -0.41 & 8.9 & $<1$ & 0.4 &  \\
\enddata
\tablenotetext{a}{We include only \mosel\ galaxies with spectroscopic redshift quality flag of $Q_z\geq2.5$ \citep[see][]{tran:15,nanayakkara:16}.}
\tablenotetext{b}{Galaxy identification numbers, observed \zfourge\ K-band magnitudes, photometric redshifts, and rest-frame \uvj\ are from \zfourge\ \citep{straatman:16}.  Uncertainties on the magnitudes and colors are $<0.01$.}
\tablenotetext{c}{We use the stellar masses from \citet{forrest:18} and the combined UV+IR star formation rates from \citet{tomczak:16}.  We recommend the reader consider a typical uncertainty of $\sim0.3$~dex for both parameters.}
\tablenotetext{d}{Effective radii are from \citet{vanderwel:12} and measured using the WFC3/F160W imaging.  Here we take the sizes reported in arcsec and convert to kpc using the angular diameter distance.}
\end{deluxetable}

\clearpage
\begin{deluxetable}{lrrrrr}
\tablecaption{\mosel: \OIII5007\AA\ Properties}\label{tab:spectra}
\tablewidth{0pt}
\tablehead{
\colhead{Field} &
\colhead{\zfourge\tablenotemark{a}}          &
\colhead{$f(5007$\AA)\tablenotemark{b}} &
\colhead{$\sigma_{\rm 1D}$\tablenotemark{b}} & 
\colhead{\intsigma\tablenotemark{b}} &
\colhead{\ewrest\tablenotemark{c}}\\
\colhead{} &
\colhead{ID} &
\colhead{$10^{-17}$~erg~s$^{-1}$~cm$^{-2}$ } & 
\colhead{\AA} & 
\colhead{\kms} &
\colhead{\AA}
}
\startdata
COSMOS      & 1877       & 2.2      $\pm$0.6      & 10.5    & 151   & 23.3     \\
COSMOS      & 4214         & 1.4      $\pm$1.1      & 3.0     & 39     &  23.0   \\
COSMOS      & 7239       & 10.1     $\pm$0.3      & 4.3     & 62      &   542 \\
COSMOS      & 9884       & 10.5     $\pm$0.6      & 9.0     & 124     & 392   \\
COSMOS      & 11063       & 17.3     $\pm$0.4      & 4.1     & 60      & 468   \\
COSMOS      & 11284       & 16.8     $\pm$0.9      & 4.9     & 68       & 262  \\
COSMOS      & 11544       & 12.3     $\pm$0.6      & 6.0     & 83       & 153  \\
COSMOS      & 12000       & 8.3      $\pm$0.4      & 8.5     & 119     & 78.0   \\
COSMOS      & 12105       & 9.6      $\pm$0.7      & 4.9     & 67       & 221  \\
COSMOS      & 12273       & 8.0      $\pm$0.5      & 7.7     & 110     & 78.6   \\
COSMOS      & 12776       & 11.5     $\pm$1.6      & 8.7     & 116    & 199    \\
COSMOS      & 12922       & 4.9      $\pm$1.1      & 10.0    & 141     & 67.6   \\
COSMOS      & 14984       &  8.1     $\pm$0.2    & 5.1     & 70      & 228  \\
COSMOS      & 15625       & 8.7      $\pm$0.3      & 4.1     & 58     & 199    \\
COSMOS      & 15636        & 4.8    $\pm$0.5      &  11.4  & 154   & 514      \\ 
COSMOS      & 16067       & 24.8     $\pm$0.4      & 6.3     & 90    & 428     \\
COSMOS      & 16325       & 6.7      $\pm$1.3      & 7.8     & 105    & 109    \\
COSMOS      & 16513       & 13.0     $\pm$0.6      & 4.3     & 58     & 289    \\
COSMOS      & 16518       & 4.3      $\pm$1.0      & 8.3     & 113    & 40.8    \\
COSMOS      & 16984       & 14.7     $\pm$0.3      & 4.4     & 60     & 405    \\
COSMOS      & 17008       & 10.6     $\pm$0.8      & 6.3     & 84     & 399    \\
COSMOS      & 17423       & 12.3     $\pm$1.1      & 6.9     & 90     & 285    \\
COSMOS      & 17909       & 22.5     $\pm$0.7      & 8.8     & 125    & 248    \\
COSMOS      & 18022       & 6.5      $\pm$0.4      & 6.2     & 83      & 282   \\
COSMOS      & 20001       & 22.0     $\pm$0.5      & 7.3     & 97    & 378     \\
CDFS        & 22136        & 2.8      $\pm$0.7      & 14.3    & 208     & 97.0   \\
CDFS        & 15782        & 2.6      $\pm$0.5      & 6.6     & 96       & 240  \\
CDFS        & 18053        & 1.0      $\pm$0.6      & 3.6     & 49      & 165   \\
CDFS        & 17189        & 2.3      $\pm$0.4      & 8.0     & 105    & 506    \\
CDFS        & 14864        & 2.7      $\pm$0.3      & 10.8    & 142   & 46.6     \\
CDFS        & 15561        & 3.2      $\pm$0.2      & 2.8     & 41      & 430   \\
\enddata
\tablenotetext{a}{Galaxy identification numbers are from \zfourge\ \citep{straatman:16}.  }
\tablenotetext{b}{\OIII5007\AA\ line-flux, line-width ($\sigma_{\rm 1D}$), and the corresponding integrated velocity dispersion (\intsigma) are measured by fitting a Gaussian to the MOSFIRE spectra (see \S\ref{sec:linefitting}). }
\tablenotetext{c}{\OIII5007\AA\ rest-frame equivalent widths are determined using the line flux measured with MOSFIRE spectroscopy and continuum flux from the best-fit FAST SED (see \S\ref{sec:line_properties}).}
\end{deluxetable}

\bibliographystyle{/Users/vy/Work/aastex/aastexv6.0/aasjournal}
\bibliography{/Users/vy/Work/aastex/tran}

\end{document}